\documentclass[
headsepline=false,
abstract=true,
numbers=noendperiod
]{scrartcl}
\usepackage[UKenglish]{babel}                         
\usepackage[utf8]{inputenc}							
\usepackage[T1]{fontenc}							
\usepackage{lmodern}                                
\usepackage{fourier}								
\usepackage[section]{placeins}					    
\usepackage{booktabs}								
\usepackage{csquotes}                               
\usepackage{mathtools}								
\usepackage{amsmath}                                
\usepackage{amssymb}								
\usepackage{dsfont}                                 
\usepackage[dvipsnames]{xcolor}
\usepackage[mathscr]{euscript}					    
\usepackage{tikz}                                   
\usetikzlibrary{decorations.pathmorphing,patterns,shapes}
\usetikzlibrary{arrows.meta}
\usepackage{xfrac}                                  
\usepackage{multirow}                               
\usepackage{caption}                                
\usepackage{tensor}                                 
\usepackage{stmaryrd}                               
\usepackage[
backend=biber,
style=numeric,
citestyle=numeric-comp,
sorting=none,
doi=false,
isbn=false,
maxbibnames=99
]{biblatex}                                         
\usepackage[pdfborder={0 0 0}]{hyperref}			
\usepackage{xurl}
\hypersetup{breaklinks=true}
\usepackage{microtype}                              
\newcommand{\e}[1]{\operatorname{e}^{#1}}

\renewcommand{\d}{\operatorname{d}\!}

\newcommand{\isl}{\mathfrak{isl}}
\newcommand{\iso}{\mathfrak{iso}}

\newcommand{\ihs}{\mathfrak{ihs}}

\newcommand{\R}{\mathds{R}}
\newcommand{\Msq}{\mathcal{M}^2}
\renewcommand{\S}{\mathscr{S}}
\newcommand{\C}{\mathcal{C}}
\newcommand{\Q}[4]{\,\tensor*[^{#1}_{#3}]{\mathbf{Q}}{^{#2}_{#4}}\,}
\renewcommand{\c}[4]{\,\tensor*[^{#1}_{#3}]{c}{^{#2}_{#4}}}
\newcommand{\J}[2]{\,\mathbf{J}^{#1}_{#2}}
\renewcommand{\P}[2]{\,\mathbf{P}^{#1}_{#2}}
\numberwithin{equation}{section}
\KOMAoptions{parskip=never}
\begin{document}
\title{Unfolded Fierz-Pauli Equations in Three-Dimensional Asymptotically Flat Spacetimes\\ \ }
\author{\textsc{Martin Ammon}\thanks{\href{mailto:martin.ammon@uni-jena.de}{martin.ammon@uni-jena.de}} \and \textsc{Michel Pannier}\thanks{\href{mailto:michel.pannier@uni-jena.de}{michel.pannier@uni-jena.de}}}
\publishers{%
\vspace{0.6cm}
\begin{minipage}[t]{0.4\linewidth}
\begin{center}
    {\footnotesize \emph{Theoretisch-Physikalisches Institut\\ Friedrich-Schiller-Universität Jena\\ Max-Wien-Platz 1, D-07743 Jena, Germany} \par}
\end{center}
\end{minipage}\hspace{0.5cm}
}%
\date{}
\maketitle
\thispagestyle{empty}
\setcounter{page}{0}
\begin{abstract}
We utilise a quotient of the universal enveloping algebra of the Poincaré algebra in three spacetime dimensions, on which we formulate a covariant constancy condition. The equations so obtained contain the Fierz-Pauli equations for non-interacting, massive higher-spin fields, and can thus be regarded as an unfolding of the Fierz-Pauli system. All fundamental fields completely decouple from each other. In the non-truncated case, the field content includes infinitely many copies of each field at fixed spin.
\end{abstract}
\newpage
\tableofcontents
\section{Introduction}
In recent years there has been revived interest in theories of massive and massless higher spins in flat spacetimes, both in the context of holography, such as in  \cite{Barnich:2012aw,Afshar:2013vka,Gonzalez:2013oaa,Grumiller:2014lna,Matulich:2014hea,Basu:2015evh,Riegler:2016hah,Sleight:2016xqq,Prohazka:2017lqb,Ammon:2017vwt,Campoleoni:2020ejn,Campoleoni:2021blr,Bekaert:2022ipg}, and from various other points of view, such as in \cite{Metsaev:2005ar,Fotopoulos:2007nm,Metsaev:2007rn,Manvelyan:2010jr,Fotopoulos:2010ay,Taronna:2011kt,Fuentealba:2015jma,Conde:2016izb,Ponomarev:2016lrm,Fredenhagen:2019lsz,Metsaev:2020gmb,Skvortsov:2020pnk,Metsaev:2022yvb,Ochirov:2022nqz}. Here, we are particularly interested in the introduction of propagating (massive) degrees of freedom to an otherwise purely topological gravitational theory in the case of three spacetime dimensions, an objective that is largely motivated by the successful utilisation of such additional fields in examples of higher-spin AdS$_3$/CFT$_2$ dualities \cite{Gaberdiel:2010pz,Gaberdiel:2011wb,Ammon:2011ua,Chang:2011mz,Gaberdiel:2012ku,Kessel:2018zqm}. Our proposal relies on the construction of the associative higher-spin algebra established in \cite{Ammon:2020fxs} that allows the introduction of massive scalar fields and, as will be shown here, massive fields up to any spin.

The attempt to formulate wave equations describing massive or massless fields of any spin dates back already to early works in the thirties of the last century \cite{majorana1932teoria,Dirac:1936tg}. It was argued by Fierz and Pauli \cite{fierz1939relativistic,Fierz:1939} that the dynamics of massive, freely propagating fields of arbitrary integer spin are captured by traceless tensor fields that obey a natural generalisation of the standard wave equation and a divergence-free condition,
\begin{align}\label{eq:Fierz-Pauli_intro}
    \left(\Box-m^2\right)\phi_{\mu_1\dots\mu_\sigma}=0\,, && \nabla^\mu\phi_{\mu\mu_2\dots\mu_\sigma}=0\,, && g^{\mu\nu}\phi_{\mu\nu\mu_3\dots\mu_\sigma}=0\,.
\end{align}
Here, $\phi_{\mu_1\dots\mu_\sigma}$ is a totally symmetric tensor of rank $\sigma$.

Despite the discovery of various no-go theorems concerning the consistent interaction of higher-spin fields \cite{Johnson:1960vt,Velo:1969bt,PhysRev.135.B1049,PhysRev.159.1251,Weinberg:1980kq,Aragone:1979hx}, a variety of working examples could be established \cite{Bengtsson:1983pd,Berends:1984wp,Bengtsson:1986kh,Metsaev:1996pd} (for more comprehensive literature references see, e.g., \cite{Bekaert:2010hw,Rahman:2012thy,Rahman:2015pzl}) and, eventually, work by Vasiliev and collaborators \cite{Fradkin:1986qy,Fradkin:1987ks,Prokushkin:1998bq} provided a consistent, fully interacting theory of massless higher-spin fields as well as their interaction with massive matter fields in the case of a negative cosmological constant.

At the heart of Vasiliev's construction lies the unfolded formulation of the equations of motion. The introduction of an infinite tower of auxiliary fields that obey first-order equations coupling them together is well suited in circumstances where one is dealing with a large symmetry group and allows one to switch on interactions in a manifestly Lorentz covariant manner. However, as of today, no fully interacting theory analogous to Vasiliev theory is known in flat spacetimes.

In the present work we focus on the case of 2+1 spacetime dimensions, which brings with it several simplifications. The purely topological nature of three-dimensional Einstein gravity allows one to recast it in form of a Chern-Simons gauge theory and provides a particularly simple introduction of massless higher-spin gauge fields, just by replacing the underlying isometry algebra by a suitable higher-spin extension. The goal is to add additional degrees of freedom in the form of massive scalar or higher-spin fields linearly coupled to the gravitational background.
\section{Higher-Spin Gravity and Scalar Fields in Flat Space}
We briefly review some important aspects of three-dimensional (higher-spin) gravity in asymptotically flat spacetimes, thereby fixing our notation. As far as the higher-spin generalisation is concerned, we will be following the proposal of \cite{Ammon:2020fxs}.
\subsection{Chern-Simons Formulation}
It is a well known fact that three-dimensional gravity can be formulated as a Chern-Simons gauge theory \cite{Witten:1988hc}. The gauge field $A$ is a flat one-form, $\d A+A\wedge A=0$, and it takes values in an isometry algebra that depends on the sign of the cosmological constant. In the case of asymptotically flat spacetimes the latter is the Poincaré algebra $\iso(2,1)\simeq\isl(2,\R)\simeq\mathfrak{sl}(2,\R)\inplus\R^3$, spanned by generators $J_m$, $P_m$ ($m\in\{0,\pm1\}$) with Lie brackets
\begin{subequations}
\begin{align}
    \left[J_m\,,J_n\right]&=(m-n)J_{m+n}\,,\\
    \left[J_m\,,P_n\right]&=(m-n)P_{m+n}\,,\\
    \left[P_m\,,P_n\right]&=0\,.
\end{align}
\end{subequations}
It is convenient to split the gauge field into spin connection and vielbein, $A=\omega+e$, the first being an element of the $\mathfrak{sl}(2,\R)$-subalgebra, $\omega=\omega^m J_m$, the latter being an element of the subalgebra of translations, $e=e^m P_m$.

A suitable framework for studying flat-space holography is provided by Einstein gravity in Bondi gauge,\footnote{For more general boundary conditions see, e.g., \cite{Grumiller:2017sjh}.} in which the most general asymptotically flat solution is given by the spin connection and vielbein gauge fields \cite{Afshar:2013bla,Afshar:2013vka,Gonzalez:2013oaa}
\begin{subequations}\label{eqs:e_omega_Einstein}
\begin{align}
    \omega&=\left(J_1-\frac{M(\phi)}{4}J_{-1}\right)d\!\phi\,,\\
    e&=\left(P_1-\frac{M(\phi)}{4}P_{-1}\right)d\!u+\frac{1}{2}P_{-1}d\!r+\left(r P_0-\frac{N(u,\phi)}{2}P_{-1}\right)d\!\phi\,,\label{eq:vielbein_classical}
\end{align}
\end{subequations}
where the mass aspect $M(\phi)$ and the angular momentum aspect $N(u,\phi)$ are related by the integrability condition $\partial_\phi M(\phi)=2\partial_u N(u,\phi)$, but otherwise arbitrary functions. These fields correspond to the metric
\begin{align}\label{eq:metric}
    d\!s^2=M(\phi)d\!u^2-2d\!u d\!r+2N(u,\phi)d\!u d\!\phi+r^2 d\!\phi^2
\end{align}
in outgoing Eddington-Finkelstein coordinates. Special cases include \cite{Barnich:2012aw,Prohazka:2017equ} Minkowski spacetime ($M(\phi)=-1$, $N(u,\phi)=0$) and flat-space cosmologies ($M(\phi)=M>0$, $N(u,\phi)=N\ne 0$).

An intriguing feature of the Chern-Simons formulation of gravity is its straightforward generalisation to higher-spin gravity, simply by letting the gauge field take values in some suitable higher-spin algebra, which is a well-known procedure both in AdS \cite{Blencowe:1988gj,Bergshoeff:1989ns,Campoleoni:2010zq,Gutperle:2011kf,Ammon:2011nk,Bunster:2014mua} and flat spacetimes \cite{Afshar:2013vka,Gonzalez:2013oaa,Riegler:2016hah,Ammon:2017vwt}. One candidate for such an algebra will be reviewed in the following subsection. The choice of that particular algebra is determined by its being the only known model that allows the introduction of matter coupling.
\subsection{Flat-Space Higher-Spin Gravity}
In a recent proposal \cite{Ammon:2020fxs} it was suggested to use a quotient of the universal enveloping algebra (UEA) of $\isl(2,\R)$ to introduce higher-spin charges to asymptotically flat spacetimes. Let us summarise the set-up.

The associative algebra $\mathcal{U}(\isl(2,\R))$ is spanned by (ordered) formal products of the generators $J_m$, $P_m$ and contains two second-order Casimir elements,
\begin{subequations}
\begin{align}
    \Msq&=P_0P_0-P_1P_{-1}\,,\\
    \S&=J_0P_0-\frac{1}{2}\left(J_1P_{-1}+J_{-1}P_1\right)\,,
\end{align}
\end{subequations}
which generate an ideal that can be quotiented out by setting them to a multiple of the unit element.\footnote{Our notation does not distinguish between the number a Casimir element is parameterised with and the actual element of the UEA.} The resulting quotient is an associative algebra, which we will denote by
\begin{align}
    \ihs(\Msq,\S)=\frac{\mathcal{U}(\isl(2,\R))}{\left\langle\Msq,\S\right\rangle}\,,
\end{align}
defines a Lie algebra by the usual identification of the commutator as Lie bracket.

In order to find a convenient basis for the algebra we classify its elements with respect to their behaviour under the adjoint action of $\mathfrak{sl}(2,\R)$-generators $J_m$. Any element that commutes with $J_1$ we call \emph{highest weight} and one easily sees that any highest-weight element can be built from powers of $J_1$, powers of $P_1$, powers of
\begin{align}
    \C\equiv J_0J_0-J_1J_{-1}+J_0\,,
\end{align}
which commutes with any $J_m$, since it is the Casimir element of the $\mathfrak{sl}(2,\R)$-subalgebra, and single factors of $J_0P_1-J_1P_0$, thus forcing the introduction of three parameters $(l,s,\xi)$ to label a complete set of highest-weight generators; we define
\begin{align}
	\Q{l}{s}{\xi}{s-1-\xi}:=\begin{cases}
        (J_1)^{l-\xi}\C^{\frac{\xi}{2}}(P_1)^{s-1-l}\,, & \xi \text{ even,}\\[0.15cm]
        (J_1)^{l-\xi}\C^{\left\lfloor\frac{\xi}{2}\right\rfloor}\left(J_0P_1-J_1P_0\right)(P_1)^{s-2-l}\,, & \xi \text{ odd.}
    \end{cases}
\end{align}
A complete set of algebra generators can then be defined by repeated adjoint action of $J_{-1}$ on highest-weight generators, thereby introducing a mode index $m$,
\begin{align}\label{def_Q}
	\Q{l}{s}{\xi}{m}:=(-1)^{s-1-\xi-m}\frac{(s+m-\xi-1)!}{(2s-2\xi-2)!}\operatorname{ad}_{J_{-1}}^{s-1-\xi-m}\left(\Q{l}{s}{\xi}{s-1-\xi}\right)\,.
\end{align}
Values of the indices $s\in\mathds{N}$, $l,\xi\in\mathds{N}_0$, $m\in\mathds{Z}$ are by construction restricted to
\begin{align}\label{eq:index_ranges}
	s\ge 1\,, && 0\le \xi\le 2\left\lfloor\frac{s-1}{2}\right\rfloor\,, && |m|\le s-1-\xi\,, && \xi\le l\le \begin{cases} s-1\,, & \xi\text{ even}\\ s-2\,, & \xi\text{ odd} \end{cases}
\end{align}
and the definition implies the standard commutation relation
\begin{align}\label{eq:adjointJ}
	\left[\Q{l}{s}{\xi}{m}\,, J_n\right]=\left(m-(s-\xi-1)n\right)\Q{l}{s}{\xi}{m+n}\,.
\end{align}
We will denote the associative product of $\ihs(\Msq,\S)$ in this basis as $\star$-product. A number of results on products and commutators of the above defined generators as well as remarks on the relation of $\ihs(\Msq,\S)$ to an \.{I}nönü-Wigner contraction from the AdS case can be found in \cite{Ammon:2020fxs}. In appendix \ref{app:subsec:product-rules} we provide the product rules that are needed for the present considerations.

Of particular interest will be the Lie-subalgebra spanned by generators of indices $l\in\{0,1\}$, $\xi=0$. Let us clear up notation by defining $\J{s}{m}:=(s-1)\Q{1}{s}{0}{m}$ and $\P{s}{m}:=\Q{0}{s}{0}{m}$. The brackets of this Lie algebra read
\begin{subequations}\label{eqs:commutators}
\begin{align}
\left[\J{s}{m}\,, \J{t}{n}\right]&=\frac{1}{2}\sum_{u=0}^{\left\lfloor\!\frac{s+t-4}{2}\!\right\rfloor}\!g^{st}_u(m,n)\J{s+t-2u-2}{m+n}+\frac{\S}{\Msq}\sum_{u=0}^{\left\lfloor\!\frac{s+t-3}{2}\!\right\rfloor}\!u g^{st}_u(m,n)\P{s+t-2u-2}{m+n}\,,\\
\left[\J{s}{m}\,, \P{t}{n}\right]&=\frac{1}{2}\sum_{u=0}^{\left\lfloor\!\frac{s+t-3}{2}\!\right\rfloor}\!g^{st}_u(m,n)\P{s+t-2u-2}{m+n}\,,\\
\left[\P{s}{m}\,, \P{t}{n}\right]&=0\,,\vphantom{\sum_{u=0}^{\left\lfloor\!\frac{s+t-3}{2}\!\right\rfloor}}
\end{align}    
\end{subequations}
with structure constants
\begin{align}
    g^{st}_u(m,n)&\equiv \frac{(-1)^u\mathcal{M}^{2u}}{4^{2u}u!}\frac{\mathcal{N}^{st}_{2u+1}(m,n)}{(s-\sfrac{3}{2})^{\underline{u}}(t-\sfrac{3}{2})^{\underline{u}}(s+t-u-\sfrac{5}{2})^{\underline{u}}}
\end{align}
and the mode functions
\begin{align}\label{eq:mode_fct}
    \mathcal{N}^{st}_u(m,n)=\sum_{k=0}^u (-1)^k \binom{u}{k}(s-1+m)^{\underline{u-k}}\,(s-1-m)^{\underline{k}}\,(t-1+n)^{\underline{k}}\,(t-1-n)^{\underline{u-k}}\,.
\end{align}
The symbol $a^{\underline{k}}=a(a-1)\dots(a-k+1)$ denotes the descending factorial. It is possible to write down a simple expression for the product of two $(l=0)$-generators; it reads
\begin{align}\label{eq:prodPP}
    \P{s}{m}\star\P{t}{n}=\sum_{u=0}^{\left\lfloor\!\frac{s+t-2}{2}\!\right\rfloor}\frac{(-1)^u\mathcal{M}^{2u}}{4^{2u}u!}\frac{\mathcal{N}^{st}_{2u}(m,n)}{(s-\sfrac{3}{2})^{\underline{u}}(t-\sfrac{3}{2})^{\underline{u}}(s+t-u-\sfrac{3}{2})^{\underline{u}}}\P{s+t-2u-1}{m+n}\,.
\end{align}

Note that, while the purely translational generators $\P{s}{m}$ span an ideal of the Lie algebra, the Lorentz-like generators $\J{s}{m}$ do not span a Lie-subalgebra unless we set $\S=0$. This case was further discussed in \cite{Campoleoni:2021blr}, there denoted $\ihs_3[\infty]$, and shown to be a realisation of the Schouten bracket of Killing tensors of arbitrary rank. Note that the authors of \cite{Campoleoni:2021blr} also discuss an alternative higher-spin Lie algebra, which can be obtained as another quotient of $\mathcal{U}(\isl(2,\R))$ by identification of $J_mP_n\sim P_mJ_n$ and $P_mP_n\sim 0$ (first discussed in \cite{Ammon:2020fxs} as ``left slice'').

With the above construction at hand it is possible to introduce higher-spin charges $Z^{(s)}(\phi)$ and $W^{(s)}(u,\phi)$, where $s\ge 2$ upon including the classical functions mass aspect $Z^{(2)}(\phi)\equiv M(\phi)$ and angular momentum aspect $W^{(2)}(u,\phi)\equiv N(u,\phi)$. In a Drinfeld-Sokolov-like gauge we may consider the connections
\begin{subequations}\label{eqs:e_omega_HS}
\begin{align}
    \omega&=\left(J_1-\frac{1}{4}\sum_{s=2}^{\infty}Z^{(s)}(\phi)\J{s}{-s+1}\right)d\!\phi\,,\\
    e&=\left(P_1-\frac{1}{4}\sum_{s=2}^{\infty}Z^{(s)}(\phi)\P{s}{-s+1}\right)d\!u+\frac{1}{2}P_{-1}d\!r+\left(r P_0-\frac{1}{2}\sum_{s=2}^{\infty}W^{(s)}(u,\phi)\P{s}{-s+1}\right)d\!\phi\,,
\end{align}
\end{subequations}
which imply vanishing curvature and torsion upon imposing $\partial_{\phi}Z^{(s)}(\phi)=2\partial_u W^{(s)}(u,\phi)$.
\subsection{Matter Coupling in (Higher-Spin) Gravity}\label{subsec:HS_grav}
In this section we intend to review and clarify a prescription that allows to introduce propagating massive matter fields to a classical or higher-spin background, which was first introduced in \cite{Ammon:2020fxs}; some of the results presented here are more general.

Introducing a master field $C$ as an $\ihs(\Msq,\S)$-valued zero-form, the proposed coupling equation takes on the form of a covariant constancy condition (see \cite{Kessel:2018zqm} for a related construction in the case of asymptotically AdS spacetimes) and reads
\begin{align}\label{eq:master_eq_left}
    \operatorname{D}C\equiv\d C+\left[\omega\,,C\right]_\star+e\star C=0\,.
\end{align}
The derivative so defined obeys $\operatorname{D}^2=0$, given flatness and vanishing torsion of the gauge fields.

\minisec{Gauge Invariance}
A necessary requirement on the above equation is its invariance under finite Poincaré transformations, i.e. under gauge transformations of the group $\mathit{ISO}(2,1)\simeq\mathit{ISL}(2,\R)$. The transformation behaviour of a Chern-Simons gauge field $A\mapsto g^{-1}Ag+g^{-1}\d g$, with $g\in\mathit{ISL}(2,\R)$, upon splitting the group element into a Lorentz and a translation part,
\begin{align}
    g=g_{\text{T}}g_{\text{L}}\,, && g_{\text{L}}=\exp\left(\xi_{\text{L}}^mJ_m\right)\equiv \e{\xi_{\text{L}}}\,,\ \ g_{\text{T}}=\exp\left(\xi_{\text{T}}^mP_m\right)\equiv \e{\xi_{\text{T}}}\,,
\end{align}
implies the following transformation behaviour of spin connection and vielbein:
\begin{subequations}
\begin{align}
    \omega &\mapsto g_{\text{L}}^{-1}\left(\omega+\d\,\right)g_{\text{L}}\,,\\
    e &\mapsto g_{\text{L}}^{-1}\left(e+\left[\omega,\xi_{\text{T}}\right]+\d\xi_{\text{T}}\right)g_{\text{L}}\,.
\end{align}
\end{subequations}
Then equation \eqref{eq:master_eq_left} remains invariant under $\mathit{ISL}(2,\R)$-transformations, if we prescribe to the master field the transformation behaviour
\begin{align}\label{eq:C_gauge_trafo_left}
    C\mapsto g_{\text{L}}^{-1}g_{\text{T}}^{-1}C g_{\text{L}}\,.
\end{align}

\minisec{Unfolded Klein-Gordon Equation}
A simple starting point for the following considerations is to restrict the master field to be purely translational, i.e. to expand
\begin{align}\label{eq:C_expansion_P}
    C=\sum_{s=1}^{\infty}\sum_{|m|\le s-1} c^s_m \P{s}{m}\,,
\end{align}
where the coefficients depend on the spacetime coordinates, $c^s_m=c^s_m(u,r,\phi)$. Using this expansion as well as the classical gauge fields \eqref{eqs:e_omega_Einstein} and the product rules \eqref{eq:prodPP}, the master equation results in the set of equations on the coefficients
\begin{subequations}\label{eqs:unfolded_KG_classical}
\begin{align}
    0&=\partial_u c^s_m+c^{s-1}_{m-1}-\frac{(s+1-m)^{\underline{2}}\Msq}{4(s+\sfrac{1}{2})^{\underline{2}}}c^{s+1}_{m-1}-\frac{M}{4}\left(c^{s-1}_{m+1}-\frac{(s+1+m)^{\underline{2}}\Msq}{4(s+\sfrac{1}{2})^{\underline{2}}}c^{s+1}_{m+1}\right)\,,\\
    0&=\partial_r c^s_m+\frac{1}{2}c^{s-1}_{m+1}-\frac{1}{2}\frac{(s+1+m)^{\underline{2}}\Msq}{4(s+\sfrac{1}{2})^{\underline{2}}}c^{s+1}_{m+1}\,,\\
\begin{split}
    0&=\partial_\phi c^s_m+(s-m)c^s_{m-1}+\frac{M}{4}(s+m)c^s_{m+1}+r\left(c^{s-1}_m+\frac{(s+m)(s-m)\Msq}{4(s+\sfrac{1}{2})^{\underline{2}}}c^{s+1}_m\right)\\
    &\quad -\frac{N}{2}\left(c^{s-1}_{m+1}-\frac{(s+1+m)^{\underline{2}}\Msq}{4(s+\sfrac{1}{2})^{\underline{2}}}c^{s+1}_{m+1}\right)\,,
\end{split}
\end{align}
\end{subequations}
where the coordinate dependence of all functions is suppressed in the notation. This set can be reduced to a second-order equation for the lowest-spin component $\phi\equiv c^1_0$, which reads
\begin{align}\label{eq:KG_Einstein}
    \left(\Box^{(0)}-\Msq\right)\phi=0\,,
\end{align}
with the operator
\begin{align}\label{eq:box0}
    \Box^{(0)}\equiv \left(-M+\frac{N^2}{r^2}\right)\partial^2_{\!r}-2\partial_{\!u}\partial_{\!r}+\frac{2N}{r^2}\partial_{\!r}\partial_{\!\phi}+\frac{\partial_{\!\phi}^2}{r^2}-\frac{\partial_{\!u}}{r}+\left(-M-\frac{N^2}{r^2}+\frac{\partial_{\!\phi}N}{r}\right)\frac{\partial_{\!r}}{r}-\frac{N}{r^3}\partial_{\!\phi}\,,
\end{align}
where $M=M(\phi)$ and $N=N(u,\phi)$, being the d'Alembert operator in the metric \eqref{eq:metric}.

This consideration shows that, first, the master equation $\eqref{eq:master_eq_left}$ indeed provides an unfolded formulation of the Klein-Gordon equation \eqref{eq:KG_Einstein} and, second, that the parametrisation of the Casimir element $\Msq$ is to be identified with the mass squared of the Klein-Gordon field (in close analogy to the AdS case \cite{Gaberdiel:2012uj}).

\minisec{Generalised Klein-Gordon Equation}
We may use the higher-spin gauge fields \eqref{eqs:e_omega_HS}, while still keeping the ansatz \eqref{eq:C_expansion_P}. First, writing out the matter-coupling equation in spacetime components on the level of the master field, we have
\begin{subequations}\label{eqs:spacetime-comps_master-field}
\begin{align}
    0&=\partial_u C+P_1\star C-\frac{1}{4}\sum_{s=2}^\infty Z^{(s)}(\phi)\P{s}{-s+1}\star C\,,\\
    0&=\partial_r C+\frac{1}{2}P_{-1}\star C\,,\\
    0&=\partial_\phi C+\left[J_1\,,C\right]+r P_0\star C-\frac{1}{4}\sum_{s=2}^\infty\left(Z^{(s)}(\phi)\left[\J{s}{-s+1}\,,C\right]_\star+2W^{(s)}(u,\phi)\P{s}{-s+1}\star C\right)\,.
\end{align}
\end{subequations}
Using the commutation relations \eqref{eqs:commutators} and the product rules \eqref{eq:prodPP} this yields the set of first-order differential equations that we noted down in \eqref{eqs:unfolded_HSgrav} of the appendix. However, a significantly easier way to proceed is to stay at the level of the master field and to try and assemble the charge-free part of the scalar Klein-Gordon operator $-2\partial_u\partial_r-\partial_u/r+\partial_\phi^2/r^2$ acting on $C$ from equations \eqref{eqs:spacetime-comps_master-field}. Comparing coefficients of the unit element, we arrive at the generalised Klein-Gordon equation
\begin{align}
    \left(\Box^{(\text{hs})}-\Msq\right)\phi=0
\end{align}
for $\phi\equiv c^1_0$ with the higher-spin Klein-Gordon operator
\begin{align}
\begin{split}
    \Box^{(\text{hs})} &\equiv \sum_{s=2}^\infty (-1)^{s-1}2^{s-2}\left(\frac{Z^{(s)}}{r}\partial_r\left(r \partial_r^{s-1}\right)-\frac{1}{r}\partial_\phi\left(W^{(s)}\partial_r^{s-1}\right)-\frac{W^{(s)}}{r}\partial_r^{s-1}\left(\frac{\partial_\phi}{r}\right)\right)\\
    &\quad +\frac{1}{r}\sum_{s,s'=2}^\infty (-1)^{s+s'}2^{s+s'-4}W^{(s)}W^{(s')}\partial_r^{s-1}\left(\frac{\partial_r^{s'-1}}{r}\right) -2\partial_u\partial_r-\frac{\partial_u}{r} +\frac{\partial_\phi^2}{r^2}\,,
\end{split}
\end{align}
where $Z^{(s)}=Z^{(s)}(\phi)$ and $W^{(s)}=W^{(s)}(u,\phi)$.

All in all, it appears reasonable to assume that the dynamics of a scalar field, both on a classical and a higher-spin background, are accurately captured by equation \eqref{eq:master_eq_left} if the master field is restricted to the $(l=0)$-part of $\ihs(\Msq,\S)$; its lowest component $c^1_0$ is to be identified with the physical scalar field $\phi$, while all higher components are auxiliary fields that can be expressed through derivatives of $c^1_0$.
\section{Massive Fields of Arbitrary Spin Unfolded}
In this section we will show how the inclusion of higher-$l$ generators into the expansion of the master field $C$ allows the description of massive fields of spin $l$. As it turns out, it will in general be possible to to truncate the master field to some finite $\ell$, thus including fields of spin $0,\,1,\,\dots,\ell$. Letting $\ell\rightarrow\infty$ one obtains the complete Fierz-Pauli system.

In this section we will use our master equation with a slight modification, where the vielbein is acting multiplicatively from the right,
\begin{align}\label{eq:master_eq_right}
    \d C+\left[\omega\,,C\right]_\star+C\star e=0\,.
\end{align}
This will not alter the physical content of the equation and is only due to the significantly easier form of the product rules $\Q{l}{s}{\xi}{m}\star P_n$ compared to left multiplication.\footnote{Note that, in the case of right multiplication, the gauge transformation behaviour \eqref{eq:C_gauge_trafo_left} of the master field must be modified to be $C\mapsto g_{\text{L}}^{-1}C g_{\text{T}}^{-1}g_{\text{L}}$.}
\subsection{Recovering the Proca Equation in Minkowski Spacetime}
To illustrate the working mechanism we start with the case of the spin-1 massive field, using Minkowski spacetime as background geometry for simplicity. We expand the master field into generators of $l=0$ and $l=1$:
\begin{align}
    C=\sum_{s=1}^\infty\sum_{|m|\le s-1} \c{0}{s}{0}{m}\Q{0}{s}{0}{m}+\sum_{s=2}^\infty\sum_{|m|\le s-1} \c{1}{s}{0}{m}\Q{1}{s}{0}{m}+\sum_{s=3}^\infty\sum_{|m|\le s-2} \c{1}{s}{1}{m}\Q{1}{s}{1}{m}\,,
\end{align}
such that the lowest components, respectively the fundamental fields, are the scalar field $\c{0}{1}{0}{0}$ at $l=0$ and $\c{1}{2}{0}{m}$ with $m\in\{0,\pm 1\}$ at $l=1$. Fields of index $s\ge 2$ at $l=0$ or $s\ge 3$ at $l=1$ as well as fields of index $\xi=1$ are generally considered auxiliary. We will argue below that this identification is justified.

The gauge background in the case of Minkowski spacetime is given by $\eqref{eqs:e_omega_Einstein}$ with $M=-1$ and $N=0$, such that, in components, equation \eqref{eq:master_eq_right} reads
\begin{subequations}
\begin{align}
    0&=\partial_u C+C\star\left(P_1+\frac{1}{4}P_{-1}\right)\,,\\
    0&=\partial_r C+\frac{1}{2}C\star P_{-1}\,,\\
    0&=\partial_\phi C+\left[J_1+\frac{1}{4}J_{-1}\,,C\right]+r C\star P_0\,.
\end{align}
\end{subequations}
Using the commutation relations \eqref{eq:adjointJ} and the product rules \eqref{eqs:QP} we can write out these expressions in algebra components and find the following set of first-order differential equations (unfolded equations):
\begingroup\allowdisplaybreaks
\begin{subequations}
\begin{align}
& \hspace{-1em}\underline{l=0,\ \xi=0:} \notag\\
    0&=\partial_u\c{0}{s}{0}{m}-\frac{1}{2}\partial_r\c{0}{s}{0}{m}+\c{0}{s-1}{0}{m-1}-\frac{(s+1-m)^{\underline{2}}}{4s^2(s+\sfrac{1}{2})^{\underline{2}}}\left(s^2\Msq\c{0}{s+1}{0}{m-1}+\S\c{1}{s+1}{0}{m-1}\right)-\frac{(s-m)\S}{s^{\underline{2}}}\c{1}{s+1}{1}{m-1}\,,\label{eq:Proca_00_u}\\
    0&=\partial_r\c{0}{s}{0}{m}+\frac{1}{2}\c{0}{s-1}{0}{m+1}-\frac{(s+1+m)^{\underline{2}}}{8s^2(s+\sfrac{1}{2})^{\underline{2}}}\left(s^2\Msq\c{0}{s+1}{0}{m+1}+\S\c{1}{s+1}{0}{m+1}\right)+\frac{(s+m)\S}{2s^{\underline{2}}}\c{1}{s+1}{1}{m+1}\,,\label{eq:Proca_00_r}\\
    \begin{split}
    0&=\partial_\phi\c{0}{s}{0}{m}+(s-m)\c{0}{s}{0}{m-1}-\frac{s+m}{4}\c{0}{s}{0}{m+1}+r\c{0}{s-1}{0}{m}\\
    &\quad +\frac{(s+m)(s-m)r}{4s^2(s+\sfrac{1}{2})^{\underline{2}}}\left(s^2\Msq\c{0}{s+1}{0}{m}+\S\c{1}{s+1}{0}{m}\right)+\frac{m r \S}{s^{\underline{2}}}\c{1}{s+1}{1}{m}\,\label{eq:Proca_00_phi};
    \end{split}\\[0.3cm]
& \hspace{-1em}\underline{l=1,\ \xi=0:} \notag\\
    0&=\partial_u\c{1}{s}{0}{m}-\frac{1}{2}\partial_r\c{1}{s}{0}{m}+\c{1}{s-1}{0}{m-1}-\frac{(s^2-1)(s+1-m)^{\underline{2}}\Msq}{4s^2(s+\sfrac{1}{2})^{\underline{2}}}\c{1}{s+1}{0}{m-1}+\frac{(s-m)\Msq}{s^{\underline{2}}}\c{1}{s+1}{1}{m-1}\,,\label{eq:Proca_10_u}\\
    0&=\partial_r\c{1}{s}{0}{m}+\frac{1}{2}\c{1}{s-1}{0}{m+1}-\frac{(s^2-1)(s+1+m)^{\underline{2}}\Msq}{8s^2(s+\sfrac{1}{2})^{\underline{2}}}\c{1}{s+1}{0}{m+1}-\frac{(s+m)\Msq}{2s^{\underline{2}}}\c{1}{s+1}{1}{m+1}\,,\label{eq:Proca_10_r}\\
    \begin{split}
    0&=\partial_\phi \c{1}{s}{0}{m}+(s-m)\c{1}{s}{0}{m-1}-\frac{s+m}{4}\c{1}{s}{0}{m+1}+r\c{1}{s-1}{0}{m}\\
    &\quad +\frac{(s^2-1)(s+m)(s-m)r\Msq}{4s^2(s+\sfrac{1}{2})^{\underline{2}}}\c{1}{s+1}{0}{m}-\frac{m r \Msq}{s^{\underline{2}}}\c{1}{s+1}{1}{m}\,\label{eq:Proca_10_phi};
    \end{split}\\[0.3cm]
& \hspace{-1em}\underline{l=1,\ \xi=1:} \notag\\
    0&=\partial_u\c{1}{s}{1}{m}-\frac{1}{2}\partial_r\c{1}{s}{1}{m}+\c{1}{s-1}{1}{m-1}-\frac{s(s-2)(s-m)^{\underline{2}}\Msq}{4(s-1)^2(s-\sfrac{1}{2})^{\underline{2}}}\c{1}{s+1}{1}{m-1}+\frac{s-1-m}{(s-1)^{\underline{2}}}\c{1}{s-1}{0}{m-1}\,,\\
    0&=\partial_r\c{1}{s}{1}{m}+\frac{1}{2}\c{1}{s-1}{1}{m+1}-\frac{s(s-2)(s+m)^{\underline{2}}\Msq}{8(s-1)^2(s-\sfrac{1}{2})^{\underline{2}}}\c{1}{s+1}{1}{m+1}-\frac{s-1+m}{2(s-1)^{\underline{2}}}\c{1}{s-1}{0}{m+1}\,,\\
    \begin{split}
    0&=\partial_\phi\c{1}{s}{1}{m}+(s-1-m)\c{1}{s}{1}{m-1}-\frac{s-1+m}{4}\c{1}{s}{1}{m+1}+r\c{1}{s-1}{1}{m}\\
    &\quad +\frac{s(s-2)(s-1+m)(s-1-m)r\Msq}{4(s-1)^2(s-\sfrac{1}{2})^{\underline{2}}}\c{1}{s+1}{1}{m}-\frac{m r}{(s-1)^{\underline{2}}}\c{1}{s-1}{0}{m}\,.\label{eq:Proca_11_phi}
    \end{split}
\end{align}
\end{subequations}
\endgroup
Note that the $(l=0)$-equations are now coupled to components of $l=1$, both through auxiliary fields and on the level of the fundamental fields. However, we can still reduce this set of equations to the Klein-Gordon equation \eqref{eq:KG_Einstein}, only containing the scalar field.

Considering the above expressions for $\c{1}{2}{0}{m}$ and $\c{1}{3}{1}{m}$, only, one may combine various equations in such a way that all higher-spin auxiliary fields get eliminated. A linear combination of the first-order equations yields
\begin{align}\label{eq:Proca_div}
    \left(\partial_u-\frac{1}{2}\partial_r-\frac{1}{2r}\right)\c{1}{2}{0}{1}+\frac{1}{r}\partial_\phi\c{1}{2}{0}{0}+2\left(\partial_r+\frac{1}{r}\right)\c{1}{2}{0}{-1}=0\,,
\end{align}
while a linear combination including derivatives of the first-order equations allows one to obtain the equations 
\begin{subequations}\label{eqs:Proca_box}
\begin{align}
    \left(\Box^{(0)}-\Msq\right)\left(\c{1}{2}{0}{1}+4\c{1}{2}{0}{-1}\right)&=0\,,\\
    \left(\Box^{(0)}-\Msq\right)\c{1}{2}{0}{0}&=\frac{1}{r^2}\partial_\phi\left(\c{1}{2}{0}{1}-4\c{1}{2}{0}{-1}\right)+\frac{1}{r^2}\c{1}{2}{0}{0}\,,\\
    \left(\Box^{(0)}-\Msq\right)\left(\c{1}{2}{0}{1}-4\c{1}{2}{0}{-1}\right)&=\frac{1}{r^2}\left(\c{1}{2}{0}{1}-4\c{1}{2}{0}{-1}\right)\,.
\end{align}
\end{subequations}
The scalar d'Alembert operator $\Box^{(0)}$ is defined in \eqref{eq:box0}. These equations are precisely the Proca equations for a spin-one field of mass $\mathcal{M}$,
\begin{align}
    \left(\Box-\Msq\right)\phi_\mu=0\,, && \nabla^\mu\phi_\mu=0\,,
\end{align}
when the field is written in the basis given by the vielbein \eqref{eq:vielbein_classical} for $M=-1$ and $N=0$, i.e.
\begin{align}
    \phi_u=-\frac{1}{2}\left(\phi^1+4\phi^{-1}\right)\,, && \phi_r=-\phi^1\,, && \phi_\phi=r\phi^0\,,
\end{align}
and, finally, using the identification $\c{1}{2}{0}{m}=\phi^m$.

To conclude this example, we have to address the question of the actual independence of fundamental fields of different spin from each other, since the appearance of higher-spin terms in lower-spin equations, particularly \eqref{eq:Proca_00_u}, \eqref{eq:Proca_00_r} and \eqref{eq:Proca_00_phi}, may lead to the suspicion that the fundamental fields are not independent after all. For example, it might be the case that the Proca field can be directly expressed in terms of derivatives of the scalar field, thus rendering it auxiliary, or that the ubiquitous auxiliary fields give rise to a non-trivial coupling in the form of higher-order differential equations. This is, however, not the case.

One may treat equations \eqref{eq:Proca_10_u} and \eqref{eq:Proca_10_r} as recurrence relations for $\c{1}{s}{0}{m}$ and $\c{1}{s}{1}{m}$ and similarly one of the equations \eqref{eq:Proca_00_u} or \eqref{eq:Proca_00_r} as recurrence relations for $\c{0}{s}{0}{m}$. Only in the case of highest- or lowest-weight components one has to resort to $\phi$-equations. This uses up a significant number of equations already. For the moment, let us argue only through explicit calculation: one may use the recurrence relations to evaluate all remaining first-order equations at lower orders of the index $s$ and find that all these are satisfied automatically, once the equations of motion \eqref{eq:Proca_div} and \eqref{eqs:Proca_box} are employed. This is not a proof, of course, but we will strengthen this observation in the general case below.
\subsection{Recovering the Fierz-Pauli Equations in Asymptotically Flat Spacetimes}
The same procedure that was used in the previous subsection can be applied in the case of a master field that is truncated to $l\le 2$, which should then result in the equations of linearised topologically massive gravity in addition to the Proca and Klein-Gordon equations. This is indeed the case, only with the peculiarity the an additional scalar field shows up in the form of $\c{2}{3}{2}{0}$, giving rise to the question whether or not the Fierz-Pauli fields here described are actually traceless. This new scalar field, however, completely decouples from the rest and could, in principle, be set to zero in the end. We will not present the details of the spin-two case but rather step forward to the general case, immediately.

The master field is expanded into the complete set of $\ihs(\Msq,\S)$-generators like
\begin{align}
    C=\sum_{\ihs} \c{l}{s}{\xi}{m}(u,r,\phi)\Q{l}{s}{\xi}{m}\,,
\end{align}
where the sum runs over all index combinations $(l,s,\xi,m)$ allowed by \eqref{eq:index_ranges}. If necessary, one may still think of the expansion as being truncated to some finite index $\ell$, which is simply achieved by setting components of $l>\ell$ to zero.

Though the lowest components at fixed $l$ are $\c{s-1}{s}{0}{m}$ with $|m|\le s-1$, there is more to take care about: our notion of spin is tied to the behaviour of algebra generators under the commutation relation \eqref{eq:adjointJ} and translates to the spin we associate to any particular field. It may therefore not come as a surprise that we actually have to take into account all fields $\c{s-1}{s}{\xi}{m}$ with $|m|\le s-\xi-1$ as being fundamental fields of spin $\sigma=s-\xi-1$, where $0\le \xi\le 2\lfloor(s-1)/2\rfloor$ and $\xi$ is an even number. That is, for any spin $\sigma$ we get an infinite number of massive fields, as long as no truncation is made.

The gauge background in the case of a general asymptotically flat spacetime is given by $\eqref{eqs:e_omega_Einstein}$, such that equation \eqref{eq:master_eq_right} reads
\begin{subequations}\label{eqs:master_components}
\begin{align}
    0&=\partial_u C+C\star\left(P_1-\frac{M}{4}P_{-1}\right)\,,\\
    0&=\partial_r C+\frac{1}{2}C\star P_{-1}\,,\\
    0&=\partial_\phi C+\left[J_1-\frac{M}{4}J_{-1}\,,C\right]+C\star\left(r P_0-\frac{N}{2}P_{-1}\right)\,.
\end{align}
\end{subequations}
Coordinate dependencies are again suppressed in the notation. Using the commutation relations \eqref{eq:adjointJ} and the product rules \eqref{eqs:QP} we get three sets of first-order differential equations on $\c{l}{s}{\xi}{m}$, which can be found in the appendix as equations \eqref{eqs:complete_first-order}, and we are interested in a linear combination of these equations that de-couples the field components $\c{s-1}{s}{\xi}{m}$ from any other fields. Such a linear combination can be found and it results in the first-order equations
\begin{align}\label{eq:div_mode_basis}
\begin{split}
    0&=\left(\partial_u+\frac{M}{2}\partial_r+\frac{M}{2r}(s-\xi-m)\right)\c{s-1}{s}{\xi}{m}+\frac{2}{r}\frac{s-\xi-m}{s-\xi+m-1}\left(\partial_\phi+N\partial_r\right)\c{s-1}{s}{\xi}{m-1}\\
    &\quad +2\frac{(s-\xi-m+1)^{\underline{2}}}{(s-\xi+m-1)^{\underline{2}}}\left(\partial_r+\frac{s-\xi+m-2}{r}\right)\c{s-1}{s}{\xi}{m-2}\,.
\end{split}
\end{align}
Here $-(s-\xi-3)\le m\le s-\xi-1$, such that there are $2s-2\xi-3$ equations at fixed $s-\xi$.

In order to find a set of second-order partial differential equations for the set of field components $\c{s-1}{s}{\xi}{m}$ it is convenient to first assemble the scalar Klein-Gordon operator $\Box^{(0)}-\Msq$ from the equations \eqref{eqs:master_components}, which results in
\begin{align}
    \left(\Box^{(0)}-\Msq\right)C=\frac{1}{r^2}\operatorname{ad}^2_{\omega_\phi}(C)+\frac{2}{r}\operatorname{ad}_{\omega_\phi}\left(C\star P_0+\frac{N}{2r^2}C\right)+\frac{\partial_\phi M}{4r^2}\left[J_{-1}\,,C\right]\,,
\end{align}
where the scalar d'Alembert operator $\Box^{(0)}$ was defined in \eqref{eq:box0}. Finally, by using the first-order equations for $\partial_r\c{s-1}{s}{\xi}{m\pm 1}$ and $\partial_\phi \c{s-1}{s}{\xi}{m\pm 1}$ to decouple $\c{s-1}{s}{\xi}{m}$ from any other fields one arrives at the set of equations
\begin{align}\label{eq:FP_mode_basis}
\begin{split}
    0&=\left(\Box^{(0)}+\frac{M}{2r^2}\left(\left(s-\xi\right)^{\underline{2}}-m^2\right)-\Msq\right)\c{s-1}{s}{\xi}{m}\\
    &\quad +\frac{2}{r^2}(s-\xi-m)\left(\partial_\phi+N\partial_r-\frac{N}{2r}\right)\c{s-1}{s}{\xi}{m-1}\\
    &\quad +\frac{M}{2r^2}(s-\xi+m)\left(\partial_\phi+N\partial_r-\frac{N}{2r}+\frac{\partial_\phi M}{2M}\right)\c{s-1}{s}{\xi}{m+1}\\
    &\quad +\frac{1}{r^2}(s-\xi-m+1)^{\underline{2}}\,\c{s-1}{s}{\xi}{m-2}+\left(\frac{M}{4r}\right)^2(s-\xi+m+1)^{\underline{2}}\,\c{s-1}{s}{\xi}{m+2}\,.
\end{split}
\end{align}
We will in the following show that equations \eqref{eq:div_mode_basis} and \eqref{eq:FP_mode_basis} are indeed the Fierz-Pauli equations \eqref{eq:Fierz-Pauli_intro} for a field of mass $\mathcal{M}$ and spin $s-\xi-1$ that freely propagates on an asymptotically flat spacetime.

Let us start by re-writing the Fierz-Pauli equations \eqref{eq:Fierz-Pauli_intro} for a field of spin $\sigma=s-1$. Changing from spacetime indices to flat indices by using the vielbein,
\begin{align}
    \phi_{\mu_1\dots\mu_{s-1}}=e^{n_1}_{\mu_1}\dots e^{n_{s-1}}_{\mu_{s-1}}\eta_{m_1n_1}\dots\eta_{m_{s-1}n_{s-1}}\phi^{m_1\dots m_{s-1}}\,,
\end{align}
equations \eqref{eq:Fierz-Pauli_intro} simply become
\begin{align}
    \left(\nabla^\mu\nabla_\mu-m^2\right)\phi^{m_1\dots m_{s-1}}=0\,, && e^\mu_m\nabla_\mu \phi^{m m_2\dots m_{s-1}}=0\,,
\end{align}
where the covariant derivative now acts on $\phi^{m_1\dots m_{s-1}}$ through the spin connection $\omega^m{}_n$ (noted down in \eqref{eq:spin_connection_expl} of the appendix). The irreducibility condition $g^{\mu\nu}\phi_{\mu\nu\mu_3\dots\mu_{s-1}}=0$ reduces to
\begin{align}
    \phi^{00m_3\dots m_{s-1}}=4\phi^{1-1m_3\dots m_{s-1}}\,.
\end{align}
Due to this condition and the fact that all fields are symmetric in their indices, we only need to consider fields of the form $\phi^{1\dots 1-1\ldots -1}$ and $\phi^{1\dots 10-1\ldots -1}$; let us introduce the short-hand
\begin{align}\label{eq:notation}
    (\pm 1)_k\equiv\underbrace{\pm 1 \ldots \pm 1}_k\,.
\end{align}
Then the action of the spin connection in the $\isl(2,\R)$-basis can then be obtained and we noted it down in section \ref{app:sec:metric_quantities} of the appendix (as well as the necessary metric quantities).

The final ingredient is the explicit identification of these fields with the fields $\c{s-1}{s}{0}{m}$. Indeed, upon setting
\begin{subequations}\label{eqs:identification_phi_c}
\begin{align}
    \phi^{(1)_{\frac{s-1+m}{2}}\,(-1)_{\frac{s-1-m}{2}}}&=\frac{(s-1-m)!(s-1+m)!}{(2s-2)!}\c{s-1}{s}{0}{m}\,,\\
    \phi^{(1)_{\frac{s-1+m}{2}}\,0\,(-1)_{\frac{s-1-m}{2}}}&=\frac{2(s-1-m)!(s-1+m)!}{(2s-2)!}\c{s-1}{s}{0}{m}
\end{align}
\end{subequations}
and $m^2=\Msq$ we find the wave equation in \eqref{eq:Fierz-Pauli_intro} to be precisely \eqref{eq:FP_mode_basis} for $\xi=0$ and the divergence equation to be \eqref{eq:div_mode_basis} for $\xi=0$. Obviously, the basis change works along the same lines if we label the spin by $\sigma=s-\xi-1$ and include this additional index in \eqref{eqs:identification_phi_c}, such that we arrive at equations \eqref{eq:FP_mode_basis} and \eqref{eq:div_mode_basis} for $\xi\ne 0$.

A few comments regarding the field content are in order. The number of additional fields introduced by even values of the index $\xi$ exactly matches the number of additional fields that would be introduced if the tracelessness-condition of the Fierz-Pauli system was not imposed. In the latter case, however, a field of spin $\sigma$ would be coupled to a field of spin $\sigma-2$, which is not the case in our system -- all fundamental fields are completely decoupled (see also the next subsection). In particular, we could set all fundamental higher-$\xi$ fields to zero without any obstacles.
\subsection{Decoupling of Fundamental Fields}
So far we did not make sure that the divergence equations \eqref{eq:div_mode_basis} and the wave equations \eqref{eq:FP_mode_basis} are the only constraints on the fundamental fields $\c{s-1}{s}{\xi}{m}$ and that there are no additional couplings between these fields, e.g. via higher-derivative equations. The rather involved form of the unfolded equations \eqref{eqs:complete_first-order}, in which fields of different $l$-slices are mixed together, makes it hard to see if there are any such constraints introduced through the interplay of auxiliary fields. We will in the following argue that this is not the case.

First, let us re-arrange the equations of motion by introduction of the inverse vielbein $e^\mu_m=\eta_{mn}g^{\mu\nu}e^n_\nu$ (see \eqref{eq:inverse_vielbein} and \eqref{eq:inverse_vielbein_derivative}) and the derivative
\begin{align}
    \operatorname{D}_m\equiv e^\mu_m\partial_\mu+\frac{\delta_{m,0}}{r}\operatorname{ad}_{\omega_\phi}\,,
\end{align}
such that the matter-coupling equations \eqref{eqs:master_components} read
\begin{align}\label{eq:master_compact}
    \operatorname{D}_m C+C\star P_m=0\,,
\end{align}
which is probably the most compact form that can be achieved. From this one may extract an equation for $\partial_\mu \c{l}{s}{\xi}{m}$.

We start at the simplest case under question, namely coupling equations of the fundamental fields up to first order in derivatives. The most general form of such an equation is
\begin{align}\label{eq:ansatz_first-order}
    \sum_{s=1}^\infty\sum_{\substack{\xi=0\\ \xi\text{ even}}}^{2\left\lfloor\frac{s-1}{2}\right\rfloor}\sum_{|m|\le s-\xi-1}\left(\alpha^\mu(s,\xi,m)\partial_\mu \c{s-1}{s}{\xi}{m}+\alpha(s,\xi,m)\c{s-1}{s}{\xi}{m}\right)=0\,.
\end{align}
Replacing the derivative therms in this equations using the first-order equations one can find conditions on the coefficients $\alpha^\mu(s,\xi,m)$ and $\alpha(s,\xi,m)$. These are
\begin{subequations}
\begin{align}
    0&=\sum_{n=-1}^1\mathcal{N}^{s-\xi,2}_1(m,n)e^n_\mu\alpha^\mu(s,\xi,m+n)\,,\\
    0&=\sum_{n=-1}^1\mathcal{N}^{s-\xi+1,2}_2(m,n)e^n_\mu\alpha^\mu(s,\xi,m+n)\,,
\end{align}
\end{subequations}
as well as $\alpha(s,\xi,m)=(s-\xi-m-1)\alpha^\phi(s,\xi,m+1)+(s-\xi+m-1)M/4\alpha^\phi(s,\xi,m-1)$. The important point is that coefficients of different parameters $s$ or $\xi$ do not mix, such that there is one set of relations for every fundamental field. Accordingly, the divergence equations we presented in \eqref{eq:div_mode_basis} are the only first-order differential equations of the general form \eqref{eq:ansatz_first-order} allowed by the unfolded equations (as long as no purely algebraic relations between fields are imposed).

We repeat the same procedure for the case of second-order differential equations. The most general expression involving only fundamental fields in this case is
\begin{align}\label{eq:ansatz_2ndOrder}
    \sum_{s=1}^\infty\sum_{\substack{\xi=0\\ \xi\text{ even}}}^{2\left\lfloor\frac{s-1}{2}\right\rfloor}\sum_{|m|\le s-\xi-1}\left(\alpha^{\mu\nu}(s,\xi,m)\partial_\mu\partial_\nu \c{s-1}{s}{\xi}{m}+\alpha^\mu(s,\xi,m)\partial_\mu \c{s-1}{s}{\xi}{m}+\alpha(s,\xi,m)\c{s-1}{s}{\xi}{m}\right)=0\,.
\end{align}
Then we get a set of recurrence relations for $\alpha^{\mu\nu}(s,\xi,m)$,
\begin{subequations}
\begin{align}
    0&=\sum_{n,n'=-1}^1\mathcal{N}_2^{s-\xi+2,2}(m,n)\mathcal{N}_2^{s-\xi+1,2}(m+n,n')e_\mu^n e_\nu^{n'} \alpha^{\mu\nu}(s,\xi,m+n+n')\,,\\\
\begin{split}
    0&=\sum_{n,n'=-1}^1\left((s-\xi+1)\mathcal{N}_1^{s-\xi,2}(m+n,n')\mathcal{N}_2^{s-\xi+1,2}(m,n)\right.\\
    &\quad\quad\quad\quad \left. +(s-\xi-1)\mathcal{N}_1^{s-\xi+1,2}(m,n)\mathcal{N}_2^{s-\xi+1,2}(m+n,n')\right)e_\mu^n e_\nu^{n'}\alpha^{\mu\nu}(s,\xi,m+n+n')\,,
\end{split}
\end{align}
\end{subequations}
as well as equations for $\alpha^\mu(s,\xi,m)$,
\begin{subequations}
\begin{align}
\begin{split}
    & \sum_{n=-1}^1\mathcal{N}_2^{s-\xi+1,2}(m,n)\left(e_\mu^n\alpha^\mu(s,\xi,m+n)+\left(\partial_\mu e_\nu^n\right)\alpha^{\mu\nu}(s,\xi,m+n)\right)\\
    & \quad\quad =\sum_{n=-1}^1 e_\mu^n \left((s-\xi-m-n-1)\mathcal{N}_2^{s-\xi+1,2}(m,n)\right.\\
    & \quad\quad\quad\quad\quad\quad\quad\quad \left.+(s-\xi-m)\mathcal{N}_2^{s-\xi+1}(m+1,n)\right)\alpha^{\mu\phi}(s,\xi,m+n+1)\\
    & \quad\quad\quad +\frac{M}{4}\sum_{n=-1}^1 e_\mu^n \left((s-\xi+m+n-1)\mathcal{N}_2^{s-\xi+1,2}(m,n)\right.\\
    & \quad\quad\quad\quad\quad\quad\quad\quad\left.+(s-\xi+m)\mathcal{N}_2^{s-\xi+1,2}(m-1,n)\right)\alpha^{\mu\phi}(s,\xi,m+n-1)\,,
\end{split}\\
\begin{split}
    & \sum_{n=-1}^1\mathcal{N}_1^{s-\xi,2}(m,n)\left(e_\mu^n\alpha^\mu(s,\xi,m+n)+\left(\partial_\mu e_\nu^n\right)\alpha^{\mu\nu}(s,\xi,m+n)\right)\\
    & \quad\quad =\sum_{n=-1}^1 e_\mu^n\left((s-\xi-m-n-1)\mathcal{N}_1^{s-\xi,2}(m,n)\right.\\
    & \quad\quad\quad\quad\quad\quad\quad\quad \left.+(s-\xi-m-1)\mathcal{N}_1^{s-\xi}(m+1,n)\right)\alpha^{\mu\phi}(s,\xi,m+n+1)\\
    & \quad\quad\quad +\frac{M}{4}\sum_{n=-1}^1 e_\mu^n\left((s-\xi+m+n-1)\mathcal{N}_1^{s-\xi,2}(m,n)\right.\\
    & \quad\quad\quad\quad\quad\quad\quad\quad \left.+(s-\xi+m-1)\mathcal{N}_1^{s-\xi,2}(m-1,n)\right)\alpha^{\mu\phi}(s,\xi,m+n-1)\,,
\end{split}
\end{align}
\end{subequations}
and for $\alpha(s,\xi,m)$,
\begin{align}
\begin{split}
    \alpha(s,\xi,m)&=(s-\xi-m-1)\alpha^\phi(s,\xi,m+1)+(s-\xi+m-1)\frac{M}{4}\alpha^\phi(s,\xi,m-1)\\
    &\quad +(s-\xi+m-1)\frac{M'}{4}\alpha^{\phi\phi}(s,\xi,m-1)-(s-\xi-m-1)^{\underline{2}}\alpha^{\phi\phi}(s,\xi,m+2)\\
    &\quad -\left((s-\xi)^{\underline{2}}-m^2\right)\frac{M}{2}\alpha^{\phi\phi}(s,\xi,m)-(s-\xi+m-1)^{\underline{2}}\left(\frac{M}{4}\right)^2\alpha^{\phi\phi}(s,\xi,m-2)\\
    &\quad +\sum_{n,n'=-1}^1 \frac{\Msq e_\mu^ne_\nu^{n'}}{4(s-\xi)^2(s-\xi+\sfrac{1}{2})}\left((s-\xi+\sfrac{1}{2})\mathcal{N}_1^{s-\xi,2}(m,n)\mathcal{N}_1^{s-\xi,2}(m+n,n')\right.\\
    &\quad\quad\quad\quad\quad\quad\quad\quad\quad\quad\quad\quad\quad\quad\quad \left.-\mathcal{N}_2^{s-\xi+1,2}(m+n,n')\right)\alpha^{\mu\nu}(s,\xi,m+n+n')\,.
\end{split}
\end{align}
These relations show that the independence of different fundamental fields $\c{s-1}{s}{\xi}{m}$ (i.e., fields of different $s$ and $\xi$) holds up to second order in derivatives as well.

In principle, one could carry on with this kind of calculation to higher-order differential equations, the repeated insertion of $\partial_\mu c$-equations, however, introduces more and more auxiliary fields, and makes the calculation increasingly involved. To convince us of the decoupling of field we instead considered the TMG case, i.e. the case in which the fundamental field components present are two scalars, $\c{0}{1}{0}{m}$ and $\c{2}{3}{2}{0}$, the Proca field, $\c{1}{2}{0}{m}$, and the TMG field $\c{2}{3}{0}{m}$. We evaluated the general ansatz of the form \eqref{eq:ansatz_2ndOrder} up to third and fourth order and found explicit solutions for the $\alpha$-coefficients using Mathematica. Indeed, there is no mixture in these cases of coefficients of different indices $s$ and $\xi$, thus showing the decoupling of the scalar, Proca and TMG fields up to fourth order in derivatives.

We speculate that a general proof of the decoupling of fundamental equations is possible on a purely algebraic level, only by utilising the equation of motion \eqref{eq:master_compact}. Since the commutators involved do not change the indices $l$, $s$ or $\xi$ of a generator, the repeated action of differential operators $\operatorname{D}_m$ can be traced back to right multiplication with translational generators, as far as operations are concerned that do change these indices. This observation may be a starting point for a general argumentation in favour of decoupling, however, we were so far not able to construct an adequate proof.
\section{Summary and Outlook}
The main objective of this work was to shed some light on the linear coupling of massive matter fields to a gravitational (higher-spin) background, here in the case of three-dimensional asymptotically flat space-times. We showed that a covariant constancy condition evaluated on a quotient of the universal enveloping algebra of the isometry algebra, here the three-dimensional Poincaré algebra, gives rise to a set of unfolded equations that can ultimately be reduced to a system of Fierz-Pauli equations for free massive higher-spin fields, all possessing the same mass. The theory contains an infinite number of copies of each higher-spin field, but no coupling between these fields is imposed.

The picture that emerges here gives an appealing interpretation to the structure of the algebra $\ihs(\Msq,\S)$, namely to the infinite wedge spanned due to the appearance of different powers of Lorentz and translation generators in the UEA. We expect that a similar result can be achieved in the case of asymptotically AdS spacetimes by considering the larger higher-spin algebra that is obtained as a quotient of $\mathcal{U}(\mathfrak{sl}(2,\R)\oplus\mathfrak{sl}(2,\R))$. Furthermore, we speculate that the algebra obtained from the additional quotient $P_mP_n\sim 0$ of $\ihs(\Msq,\S)$ plays a role in the case of massless higher-spin fields.

While in the first part of the paper we reviewed and generalised a possible formulation of higher-spin gravity, i.e. the introduction of an infinite tower of massless higher-spin gauge fields in the Chern-Simons formalism together with suitable boundary conditions, in the second part we discussed the propagation of massive higher-spin fields in a classical, i.e. spin 2, background, only. Though we expect the basic idea to remain valid in the case of a higher-spin gauge background, explicit calculations would require full knowledge of the structure constants of the associative product of $\ihs(\Msq,\S)$, which are currently not at our disposal. It is however clear that no truncation in the spin of the massive fields will be possible anymore, since the commutator of the matter field with a higher-spin deformed spin connection will produce higher-$l$ terms.

The proposal here described is so far free of any interactions, be it of the massive higher-spin fields amongst each other or a back-reaction to the gauge background, which, in a sense, still renders the theory trivial in the sense that no known no-go theorems apply. A natural next step that will be undertaken is the introduction of interactions via suitable gauge potentials. Possibly, the infinite number of copies of higher-spin fields we found in our approach may play a role there. In this context it would also be useful to find connections to other unfolding proposals, such as \cite{Boulanger:2014vya}. In a broader context, our proposal may be viewed as a necessary first step in a bottom-up approach to the construction of a flat-space counterpart to Vasiliev theory; if existing, such a model should certainly contain the unfolded equations here presented as linearised version.

As an application of our approach in the realm of holography it could be interesting to try and derive three-point correlation functions involving one scalar or arbitrary-spin current on the dual field-theory side (which is expected to be a Carrollian field theory).

A further interesting question concerns the applicability of the method here described to 3+1 or, more general, $d+1$ spacetime dimensions. Though the higher-dimensional gravitational theories are not topological theories anymore, it should still be possible to implement (linear) matter coupling through a suitable covariant-constancy condition on the universal enveloping algebra of the respective Poincaré algebra.
\section*{Acknowledgements}
The authors would like to thank Xavier Bekaert, Nicolas Boulanger, Andrea Campoleoni, Stefan Fredenhagen, Daniel Grumiller, Simon Pekar, Max Riegler, and Evgeny Skvortsov for stimulating and enlightening discussions; MP expresses special thanks for the hospitality received during an inspiring stay at UMONS as well as at the University of Vienna and TU Vienna.

MA is funded by the Deutsche Forschungsgemeinschaft (DFG, German Research  Foundation) under Grant No.\,406235073 within the Heisenberg program.

MP acknowledges support by the Deutsche Forschungsgemeinschaft (DFG) under Grant No.\,406116891 within the Research Training Group RTG\,2522/1.
%
%
%
%
%
%
%
%
%
%
\appendix
\section{Metric Quantities and Spin Connection}\label{app:sec:metric_quantities}
In the case of asymptotically flat spacetimes the metric and its inverse read
\begin{align}
    \left(g_{\mu\nu}\right)=\begin{pmatrix} M(\phi) & -1 & N(u,\phi)\\ -1 & 0 & 0\\ N(u,\phi) & 0 & r^2\end{pmatrix}\,, && \left(g^{\mu\nu}\right)=\begin{pmatrix} 0 & -1 & 0\\ -1 & -M(\phi)+\frac{N(u,\phi)^2}{r^2} & \frac{N(u,\phi)}{r^2}\\ 0 & \frac{N(u,\phi)}{r^2} & \frac{1}{r^2}\end{pmatrix}\,.
\end{align}
The non-vanishing Christoffel symbols are
\begin{align}
\begin{split}
    & \Gamma^u_{\phi\phi}=r\,,\ \ \ \ \ \ \Gamma^r_{u\phi}=\Gamma^r_{\phi u}=-\frac{M'(\phi)}{2}\,,\ \ \ \ \ \ \Gamma^r_{r\phi}=\Gamma^r_{\phi r}=\frac{N(u,\phi)}{r}\,,\\
    & \Gamma^r_{\phi\phi}=r M(\phi)-\frac{N(u,\phi)^2}{r}-\partial_\phi N(u,\phi)\,,\ \ \ \ \ \ \Gamma^\phi_{r\phi}=\Gamma^\phi_{\phi r}=\frac{1}{r}\,, \ \ \ \ \ \ \Gamma^\phi_{\phi\phi}=-\frac{N(u,\phi)}{r}\,.
\end{split}
\end{align}
In $\isl(2,\R)$-components we have for the spin connection $\omega_{mn}=-\varepsilon_{mnk}\omega^k$, where $\varepsilon_{-101}=1$. Accordingly, for asymptotically flat spacetimes
\begin{align}\label{eq:spin_connection_expl}
    \omega_\phi{}^1{}_{-1}=0=\omega_\phi{}^{-1}{}_1\,, && \omega_\phi{}^1{}_0=1\,, && \omega_\phi{}^0{}_1=\frac{M(\phi)}{2}\,, && \omega_\phi{}^{-1}{}_0=\frac{M(\phi)}{4}\,, && \omega_\phi{}^0{}_{-1}=2\,.
\end{align}
Indices are moved using the form $\eta_{mn}=(-1)^m(1+m)!(1-m)!\,\delta_{m+n,0}$. The action of the spin connection in the $\isl(2,\R)$-basis for the relevant cases of ordered indices with at most one zero can be determined as
\begingroup\allowdisplaybreaks
\begin{subequations}
\begin{align}
& \hspace{-3.2cm}\underline{(m_1,\dots,m_{s-1})=\left((1)_{\frac{s-1+m}{2}},(-1)_{\frac{s-1-m}{2}}\right):} \notag\\[0.2cm]
    \left(\omega\cdot\phi\right)^{m_1\dots m_{s-1}}&=\frac{s-1+m}{2}\phi^{(1)_{\frac{s-3+m}{2}}0(-1)_{\frac{s-1-m}{2}}}+\frac{M}{4}\frac{s-1-m}{2}\phi^{(1)_{\frac{s-1+m}{2}}0(-1)_{\frac{s-3-m}{2}}}\,,\\
    \begin{split}
    \left(\omega\cdot\left(\omega\cdot\phi\right)\right)^{m_1\dots m_{s-1}}&=(s-1+m)^{\underline{2}}\,\phi^{(1)_{\frac{s-3+m}{2}}(-1)_{\frac{s+1-m}{2}}}+\frac{M}{2}\left(s^{\underline{2}}-m^2\right)\phi^{(1)_{\frac{s-1+m}{2}}(-1)_{\frac{s-1-m}{2}}}\\
    &\quad +\left(\frac{M}{4}\right)^2(s-1-m)^{\underline{2}}\,\phi^{(1)_{\frac{s+1+m}{2}}(-1)_{\frac{s-3-m}{2}}}\,;
    \end{split}\\[0.3cm]
& \hspace{-3.2cm}\underline{(m_1,\dots,m_{s-1})=\left((1)_{\frac{s-2+m}{2}},0,(-1)_{\frac{s-2-m}{2}}\right):} \notag\\[0.2cm]
    \left(\omega\cdot\phi\right)^{m_1\dots m_{s-1}}&=2(s-1+m)\phi^{(1)_{\frac{s-2+m}{2}}(-1)_{\frac{s-m}{2}}}+\frac{M}{2}(s-1-m)\phi^{(1)_{\frac{s+m}{2}}(-1)_{\frac{s-2-m}{2}}}\,,\\
    \begin{split}
    \left(\omega\cdot\left(\omega\cdot\phi\right)\right)^{m_1\dots m_{s-1}}&=(s-1+m)^{\underline{2}}\,\phi^{(1)_{\frac{s-4+m}{2}}0(-1)_{\frac{s-m}{2}}}+\frac{M}{2}\left(s^{\underline{2}}-m^2\right)\phi^{(1)_{\frac{s-2+m}{2}}0(-1)_{\frac{s-2-m}{2}}}\\
    &\quad +\left(\frac{M}{4}\right)^2(s-1-m)^{\underline{2}}\,\phi^{(1)_{\frac{s+m}{2}}0(-1)_{\frac{s-4-m}{2}}}\,.
    \end{split}
\end{align}
\end{subequations}
\endgroup

Let us also give the inverse components of the vielbein, $e^\mu_m=\eta_{mn}g^{\mu\nu}e^n_\nu$, which are
\begin{align}\label{eq:inverse_vielbein}
    e=e^\mu_mP^m\partial_\mu=P^1\partial_u+\left(\frac{M}{2}P^1+\frac{N}{r}P_0+2P^{-1}\right)\partial_r+\frac{1}{r}P^0\partial_\phi\,.
\end{align}
Accordingly,
\begin{align}\label{eq:inverse_vielbein_derivative}
    e_1^\mu\partial_\mu=\partial_u+\frac{M}{2}\partial_r\,, && e_0^\mu=\frac{1}{r}\left(N\partial_r+\partial_\phi\right)\,, && e_{-1}^\mu\partial_\mu=2\partial_r\,.
\end{align}
\section{Collection of Lengthy Expressions}\label{sec:app_equations}
In order to spare the reader the trouble of leafing through a whole lot of less handsome equations but still provide a complete communication of our results, we decided to put some of these lengthy expressions into this appendix.
\begingroup
\allowdisplaybreaks
\subsection{Product Rules of the Higher-Spin Algebra}\label{app:subsec:product-rules}
The calculations in the main part of this paper require some knowledge about the $\ihs(\Msq,\S)$-product rules. In particular, one needs the following spin-$s$-spin-2 products:
\begin{align}
    \begin{split}\label{eqs:QP}
	\Q{l}{s}{\xi}{m}\star P_{n}&=\Q{l}{s+1}{\xi}{m+n}-\frac{2\mathcal{N}^{s-\xi,2}_1(m,n)}{(s-\xi)^{\underline{2}}}\Bigg[(l-\xi)\Q{l}{s+1}{\xi+1}{m+n}-\frac{\left(l-2\left\lfloor\sfrac{\xi}{2}\right\rfloor\right)^{\underline{2}}}{2}\Q{l-1}{s}{\xi}{m+n}\\
	&\quad +(\xi-2\lfloor\sfrac{\xi}{2}\rfloor)\left((l-\xi+1)\Msq\Q{l}{s-1}{\xi-1}{m+n}-2(l-\xi+\sfrac{1}{2})\S\Q{l-1}{s-1}{\xi-1}{m+n}\right)\Bigg]\\
	&\quad +\frac{\mathcal{N}^{s-\xi,2}_2(m,n)}{2(s-\xi-1)^2(s-\xi-\sfrac{1}{2})^{\underline{2}}}\Bigg[(l-\xi)^{\underline{2}}\Q{l}{s+1}{\xi+2}{m+n}-(l-\xi)^{\underline{2}}(l-\xi-\sfrac{1}{2})\Q{l-1}{s}{\xi+1}{m+n}\\
    &\quad -\left((s-\xi-1)^2-(l-2\lfloor\sfrac{\xi}{2}\rfloor)^2\right)\Msq\Q{l}{s-1}{\xi}{m+n}-2(l-\xi)(l-2\lfloor\sfrac{\xi}{2}\rfloor-\sfrac{1}{2})\S\Q{l-1}{s-1}{\xi}{m+n} \vphantom{\frac{\left((x_1\right)^{\underline{3}}}{2}}\\
    &\quad +\frac{(l-2\lfloor\sfrac{\xi}{2}\rfloor-1)(l-2\lfloor\sfrac{\xi}{2}\rfloor)^{\underline{3}}}{4}\Q{l-2}{s-1}{\xi}{m+n}\\
    &\quad -(\xi-2\lfloor\sfrac{\xi}{2}\rfloor)\Big((l-\xi+1)^{\underline{2}}(l-\xi+\sfrac{1}{2})\Msq\Q{l-1}{s-2}{\xi-1}{m+n}\\
    &\quad -2(l-\xi)\left((l-\xi)^2-\sfrac{1}{2}\right)\S\Q{l-2}{s-2}{\xi-1}{m+n}\Big)\Bigg]\,.
\end{split}
\end{align}
The definition of the mode functions $\mathcal{N}^{st}_u(m,n)$ is equation \eqref{eq:mode_fct} of the main part and the special cases $u=1$ and $u=2$ with $t=2$ appearing here read
\begin{subequations}
\begin{align}
    \mathcal{N}^{s2}_1(m,n)&=2(m-(s-1)n)\,,\\
    \mathcal{N}^{s2}_2(m,n)&=2\left(m^2+2(s-1)(s-\sfrac{3}{2})n^2-2(s-\sfrac{3}{2})mn-(s-1)^2\right)\,.
\end{align}
\end{subequations}
The product rules were taken from \cite{Ammon:2020fxs}, where a couple of further identities can be found.
\subsection{Unfolded Scalar Field}
The case of a scalar field of mass $\mathcal{M}$ propagating in a higher-spin deformed gauge background, where $Z^{(s)}=Z^{(s)}(\phi)$ and $W^{(s)}=W^{(s)}(u,\phi)$ with $s\ge 2$ are higher-spin charges (generalising the classical charges $M(\phi)\equiv Z^{(2)}(\phi)$ and $N(u,\phi)\equiv W^{(2)}(u,\phi)$) is discussed in section \ref{subsec:HS_grav}. We derived the following first-order differential equations for components of the master field:
\begin{subequations}\label{eqs:unfolded_HSgrav}
{\scriptsize
\begin{align}
\begin{split}
    0&=\partial_u c^s_m+c^{s-1}_{m-1}-\frac{(s-m+1)^{\underline{2}}\Msq}{4(s+\sfrac{1}{2})^{\underline{2}}}c^{s+1}_{m-1}\\
    &\quad -\sum_{s'=1}^\infty\sum_{\substack{m'=s+m-s'\\ s+s'+m+m'\text{ ev.}}}^{s'-1}\frac{(-1)^{\frac{s'-s+m'-m}{2}}(m'-m)^{\underline{\frac{s'-s+m'-m}{2}}}(s'+m'-1)^{\underline{s'-s+m'-m}}\mathcal{M}^{s'-s+m'-m}Z^{(m'-m+1)}}{2^{s'-s+m'-m+2}\left(\frac{s'-s+m'-m}{2}\right)!(s'-\sfrac{3}{2})^{\underline{\frac{s'-s+m'-m}{2}}}\left(\frac{s'+s+m'-m-1}{2}\right)^{\underline{\frac{s'-s+m'-m}{2}}}}c^{s'}_{m'}\,,
\end{split}\\
    0&=\partial_r c^s_m+\frac{1}{2}c^{s-1}_{m+1}-\frac{(s+m+1)^{\underline{2}}\Msq}{8(s+\sfrac{1}{2})^{\underline{2}}}c^{s+1}_{m+1}\,,\\
\begin{split}
    0&=\partial_\phi c^s_m+(s-m)c^s_{m-1}+r c^{s-1}_m+\frac{r(s+m)(s-m)\Msq}{4(s+\sfrac{1}{2})^{\underline{2}}}c^{s+1}_m\\
    &\quad +\sum_{s'=1}^\infty\sum_{\substack{m'=s+m-s'+1\\ s+s'+m+m'\text{ odd}}}^{s'-1}\frac{(-1)^{\frac{s'-s+m'-m-1}{2}}(m'-m)^{\underline{\frac{s'-s+m'-m+1}{2}}}(s'+m'-1)^{\underline{s'-s+m'-m}}\mathcal{M}^{s'-s+m'-m-1}Z^{(m'-m+1)}}{2^{s'-s+m'-m+1}\left(\frac{s'-s+m'-m-1}{2}\right)!(s'-\sfrac{3}{2})^{\underline{\frac{s'-s+m'-m-1}{2}}}\left(\frac{s'+s+m'-m}{2}-1\right)^{\underline{\frac{s'-s+m'-m-1}{2}}}}c^{s'}_{m'}\\
    &\quad -\sum_{s'=1}^\infty\sum_{\substack{m'=s+m-s'\\ s+s'+m+m'\text{ ev.}}}^{s'-1}\frac{(-1)^{\frac{s'-s+m'-m}{2}}(m'-m)^{\underline{\frac{s'-s+m'-m}{2}}}(s'+m'-1)^{\underline{s'-s+m'-m}}\mathcal{M}^{s'-s+m'-m}W^{(m'-m+1)}}{2^{s'-s+m'-m+1}\left(\frac{s'-s+m'-m}{2}\right)!(s'-\sfrac{3}{2})^{\underline{\frac{s'-s+m'-m}{2}}}\left(\frac{s'+s+m'-m-1}{2}\right)^{\underline{\frac{s'-s+m'-m}{2}}}}c^{s'}_{m'}\,.
\end{split}
\end{align}
}%
\end{subequations}
Non-existing index combinations in $c^s_m$ or the charges are to be identified with zeros. We call $c\equiv c^1_0$. For $s=1$, $m=0$ we get
\begin{subequations}
\begin{align}
    0&=\partial_u c-\frac{2\Msq}{3}c^2_{-1}+\frac{1}{4}\sum_{s=2}^\infty\frac{(-1)^s(s-1)!\mathcal{M}^{2(s-1)}Z^{(s)}}{(s-\sfrac{1}{2})^{\underline{s-1}}}c^s_{s-1}\,,\\
    0&=\partial_r c-\frac{\Msq}{3}c^2_1\,,\\
    0&=\partial_\phi c+\frac{r\Msq}{3}c^2_0+\frac{1}{2}\sum_{s=2}^\infty\frac{(-1)^s(s-1)!\mathcal{M}^{2(s-1)}W^{(s)}}{(s-\sfrac{1}{2})^{\underline{s-1}}}c^s_{s-1}\,.
\end{align}
\end{subequations}
For the $n$-th derivative w.r.t. $r$ we find
\begin{subequations}
\begin{align}
    \partial_r^n c^s_{s-1}&=\frac{(s+n-1)^{\underline{n}}\mathcal{M}^{2n}}{2^n(s+n-\sfrac{1}{2})^{\underline{n}}}c^{s+n}_{s+n-1}\,,\\
    \partial_r^n c^s_{s-2}&=\frac{(s+n-2)^{\underline{n}}\mathcal{M}^{2n}}{2^n(s+n-\sfrac{1}{2})^{\underline{n}}}c^{s+n}_{s+n-2}\,.
\end{align}
\end{subequations}
\subsection{Unfolded Massive Equations}
Rearranging the master equation \eqref{eq:master_eq_right} in the background \eqref{eqs:e_omega_Einstein} into the form given in \eqref{eq:master_compact}, one can write down the probably most compact form of the equations of motion, namely
\begin{align}\label{eqs:complete_first-order}
\begin{split}
    0&=e^\mu_n\partial_\mu \c{l}{s}{\xi}{m+n}+\frac{\delta_{n,0}}{r}\left((s-\xi-m-n)\c{l}{s}{\xi}{m+n-1}+(s-\xi+m+n)\frac{M}{4}\c{l}{s}{\xi}{m+n+1}\right)+\c{l}{s-1}{\xi}{m}\\
    &\quad -\frac{2\mathcal{N}^{s-\xi,2}_1(m,n)}{(s-\xi)^{\underline{2}}}\Bigg[(l-\xi+1)\c{l}{s-1}{\xi-1}{m}-\frac{(l-2\lfloor\sfrac{\xi}{2}\rfloor+1)^{\underline{2}}}{2}\c{l+1}{s}{\xi}{m}\\
    &\quad +(\xi-2\lfloor\sfrac{\xi-1}{2}\rfloor+1)\left((l-\xi)\Msq\c{l}{s+1}{\xi+1}{m}-2(l-\xi+\sfrac{1}{2}_\S\c{l+1}{s+1}{\xi+1}{m}\right)\Bigg]\\
    &\quad +\frac{\mathcal{N}^{s-\xi+1,2}_2(m,n)}{2(s-\xi)^2(s-\xi+\sfrac{1}{2})^{\underline{2}}}\Bigg[(l-\xi+2)^{\underline{2}}\c{l}{s-1}{\xi-2}{m}-(l-\xi+2)^{\underline{2}}(l-\xi+\sfrac{3}{2})\c{l+1}{s}{\xi+1}{m}\\
    &\quad -\left((s-\xi)^2-(l-2\lfloor\sfrac{\xi}{2}\rfloor)^2\right)\Msq\c{l}{s+1}{\xi}{m}-2(l-\xi+1)(l-2\lfloor\sfrac{\xi}{2}\rfloor+\sfrac{1}{2})\S\c{l+1}{s+1}{\xi}{m}\\
    &\quad +\frac{(l-2\lfloor\sfrac{\xi}{2}\rfloor+1)(l-2\lfloor\sfrac{\xi}{2}\rfloor+2)^{\underline{3}}}{4}\c{l+2}{s+1}{\xi}{m}\\
    &\quad -(\xi-2\lfloor\sfrac{\xi-1}{2}\rfloor+1)\left((l-\xi+1)^{\underline{2}}(l-\xi+\sfrac{1}{2})\Msq\c{l+1}{s+2}{\xi+1}{m}\right.\\
    &\quad \left. -2(l-\xi+1)\left((l-\xi+1)^2-\sfrac{1}{2}\right)\S\c{l+2}{s+2}{\xi+1}{m}\right)\Bigg]\,.
\end{split}
\end{align}
\endgroup
\printbibliography

@article{Afshar:2013bla,
    author = "Afshar, Hamid R.",
    title = "{Flat/AdS boundary conditions in three dimensional conformal gravity}",
    eprint = "1307.4855",
    archivePrefix = "arXiv",
    primaryClass = "hep-th",
    doi = "10.1007/JHEP10(2013)027",
    journal = "JHEP",
    volume = "10",
    pages = "027",
    year = "2013"
}

@article{Prokushkin:1998bq,
    author = "Prokushkin, S. F. and Vasiliev, Mikhail A.",
    title = "{Higher spin gauge interactions for massive matter fields in 3-D AdS space-time}",
    eprint = "hep-th/9806236",
    archivePrefix = "arXiv",
    reportNumber = "FIAN-TD-16-98",
    doi = "10.1016/S0550-3213(98)00839-6",
    journal = "Nucl. Phys. B",
    volume = "545",
    pages = "385",
    year = "1999"
}

@article{Fradkin:1986qy,
    author = "Fradkin, E. S. and Vasiliev, Mikhail A.",
    title = "{Cubic Interaction in Extended Theories of Massless Higher Spin Fields}",
    reportNumber = "LEBEDEV-86-309",
    doi = "10.1016/0550-3213(87)90469-X",
    journal = "Nucl. Phys. B",
    volume = "291",
    pages = "141--171",
    year = "1987"
}

@article{Fradkin:1987ks,
    author = "Fradkin, E. S. and Vasiliev, Mikhail A.",
    title = "{On the Gravitational Interaction of Massless Higher Spin Fields}",
    doi = "10.1016/0370-2693(87)91275-5",
    journal = "Phys. Lett. B",
    volume = "189",
    pages = "89--95",
    year = "1987"
}

@article{Metsaev:2005ar,
    author = "Metsaev, R. R.",
    title = "{Cubic interaction vertices of massive and massless higher spin fields}",
    eprint = "hep-th/0512342",
    archivePrefix = "arXiv",
    reportNumber = "FIAN-TD-19-05",
    doi = "10.1016/j.nuclphysb.2006.10.002",
    journal = "Nucl. Phys. B",
    volume = "759",
    pages = "147--201",
    year = "2006"
}

@article{Berends:1984wp,
    author = "Berends, Frits A. and Burgers, G. J. H. and Van Dam, H.",
    title = "{ON SPIN THREE SELFINTERACTIONS}",
    doi = "10.1007/BF01410362",
    journal = "Z. Phys. C",
    volume = "24",
    pages = "247--254",
    year = "1984"
}

@article{Aragone:1979hx,
    author = "Aragone, C. and Deser, Stanley",
    title = "{Consistency Problems of Hypergravity}",
    reportNumber = "Print-79-0377 (BRANDEIS)",
    doi = "10.1016/0370-2693(79)90808-6",
    journal = "Phys. Lett. B",
    volume = "86",
    pages = "161--163",
    year = "1979"
}

@article{Velo:1969bt,
    author = "Velo, Giorgio and Zwanziger, Daniel",
    title = "{Propagation and quantization of Rarita-Schwinger waves in an external electromagnetic potential}",
    doi = "10.1103/PhysRev.186.1337",
    journal = "Phys. Rev.",
    volume = "186",
    pages = "1337--1341",
    year = "1969"
}

@article{Johnson:1960vt,
    author = "Johnson, Kenneth and Sudarshan, E. C. G.",
    title = "{Inconsistency of the local field theory of charged spin 3/2 particles}",
    doi = "10.1016/0003-4916(61)90030-6",
    journal = "Annals Phys.",
    volume = "13",
    pages = "126--145",
    year = "1961"
}

@article{majorana1932teoria,
  title={Teoria Relativistica di Particelle Con Momento Intrinseco Arbitrario},
  author={Majorana, Ettore},
  journal={Il Nuovo Cimento (1924-1942)},
  volume={9},
  number={10},
  pages={335--344},
  year={1932},
  publisher={Springer}
}

@ARTICLE{Fierz:1939,
       author = {{Fierz}, Markus},
        title = "{{\"U}ber die relativistische Theorie kr{\"a}ftefreier Teilchen mit beliebigem Spin}",
      journal = {Helvetica Physica Acta},
         year = 1939,
        month = jan,
       volume = {12},
        pages = {3-37},
       adsurl = {https://ui.adsabs.harvard.edu/abs/1939AcHPh..12....3F}
}

@article{fierz1939relativistic,
  title={On relativistic wave equations for particles of arbitrary spin in an electromagnetic field},
  author={Fierz, Markus and Pauli, Wolfgang Ernst},
  journal={Proceedings of the Royal Society of London. Series A. Mathematical and Physical Sciences},
  volume={173},
  number={953},
  pages={211--232},
  year={1939},
  publisher={The Royal Society London}
}

@article{Ochirov:2022nqz,
    author = "Ochirov, Alexander and Skvortsov, Evgeny",
    title = "{Chiral approach to massive higher spins}",
    eprint = "2207.14597",
    archivePrefix = "arXiv",
    primaryClass = "hep-th",
    month = "7",
    year = "2022"
}

@article{Boulanger:2014vya,
    author = "Boulanger, Nicolas and Ponomarev, Dmitry and Sezgin, Ergin and Sundell, Per",
    title = "{New unfolded higher spin systems in $AdS_3$}",
    eprint = "1412.8209",
    archivePrefix = "arXiv",
    primaryClass = "hep-th",
    reportNumber = "LMU-ASC-73-14, MIFPA-14-40",
    doi = "10.1088/0264-9381/32/15/155002",
    journal = "Class. Quant. Grav.",
    volume = "32",
    number = "15",
    pages = "155002",
    year = "2015"
}

@article{Kessel:2018zqm,
    author = "Kessel, Pan and Raeymaekers, Joris",
    title = "{Simple unfolded equations for massive higher spins in AdS$_{3}$}",
    eprint = "1805.07279",
    archivePrefix = "arXiv",
    primaryClass = "hep-th",
    doi = "10.1007/JHEP08(2018)076",
    journal = "JHEP",
    volume = "08",
    pages = "076",
    year = "2018"
}

@article{Campoleoni:2021blr,
    author = "Campoleoni, Andrea and Pekar, Simon",
    title = "{Carrollian and Galilean conformal higher-spin algebras in any dimensions}",
    eprint = "2110.07794",
    archivePrefix = "arXiv",
    primaryClass = "hep-th",
    doi = "10.1007/JHEP02(2022)150",
    journal = "JHEP",
    volume = "02",
    pages = "150",
    year = "2022"
}

@article{Ammon:2020fxs,
    author = "Ammon, Martin and Pannier, Michel and Riegler, Max",
    title = "{Scalar Fields in 3D Asymptotically Flat Higher-Spin Gravity}",
    eprint = "2009.14210",
    archivePrefix = "arXiv",
    primaryClass = "hep-th",
    doi = "10.1088/1751-8121/abdbc6",
    journal = "J. Phys. A",
    volume = "54",
    number = "10",
    pages = "105401",
    year = "2021"
}

@article{Fredenhagen:2019lsz,
    author = {Fredenhagen, Stefan and Kr\"uger, Olaf and Mkrtchyan, Karapet},
    title = "{Restrictions for $n$-Point Vertices in Higher-Spin Theories}",
    eprint = "1912.13476",
    archivePrefix = "arXiv",
    primaryClass = "hep-th",
    doi = "10.1007/JHEP06(2020)118",
    journal = "JHEP",
    volume = "06",
    pages = "118",
    year = "2020"
}

@article{Conde:2016izb,
    author = "Conde, Eduardo and Joung, Euihun and Mkrtchyan, Karapet",
    title = "{Spinor-Helicity Three-Point Amplitudes from Local Cubic Interactions}",
    eprint = "1605.07402",
    archivePrefix = "arXiv",
    primaryClass = "hep-th",
    doi = "10.1007/JHEP08(2016)040",
    journal = "JHEP",
    volume = "08",
    pages = "040",
    year = "2016"
}

@article{Manvelyan:2010jr,
    author = "Manvelyan, Ruben and Mkrtchyan, Karapet and Ruhl, Werner",
    title = "{General trilinear interaction for arbitrary even higher spin gauge fields}",
    eprint = "1003.2877",
    archivePrefix = "arXiv",
    primaryClass = "hep-th",
    doi = "10.1016/j.nuclphysb.2010.04.019",
    journal = "Nucl. Phys. B",
    volume = "836",
    pages = "204--221",
    year = "2010"
}

@article{Fuentealba:2015jma,
    author = "Fuentealba, Oscar and Matulich, Javier and Troncoso, Ricardo",
    title = "{Extension of the Poincar\'e group with half-integer spin generators: hypergravity and beyond}",
    eprint = "1505.06173",
    archivePrefix = "arXiv",
    primaryClass = "hep-th",
    reportNumber = "CECS-PHY-15-02",
    doi = "10.1007/JHEP09(2015)003",
    journal = "JHEP",
    volume = "09",
    pages = "003",
    year = "2015"
}

@article{Fotopoulos:2007nm,
    author = "Fotopoulos, Angelos and Tsulaia, Mirian",
    title = "{Interacting higher spins and the high energy limit of the bosonic string}",
    eprint = "0705.2939",
    archivePrefix = "arXiv",
    primaryClass = "hep-th",
    reportNumber = "DFTT-8-2007",
    doi = "10.1103/PhysRevD.76.025014",
    journal = "Phys. Rev. D",
    volume = "76",
    pages = "025014",
    year = "2007"
}

@article{Fotopoulos:2010ay,
    author = "Fotopoulos, A. and Tsulaia, Mirian",
    title = "{On the Tensionless Limit of String theory, Off - Shell Higher Spin Interaction Vertices and BCFW Recursion Relations}",
    eprint = "1009.0727",
    archivePrefix = "arXiv",
    primaryClass = "hep-th",
    doi = "10.1007/JHEP11(2010)086",
    journal = "JHEP",
    volume = "11",
    pages = "086",
    year = "2010"
}

@article{Matulich:2014hea,
    author = "Matulich, Javier and Perez, Alfredo and Tempo, David and Troncoso, Ricardo",
    title = "{Higher spin extension of cosmological spacetimes in 3D: asymptotically flat behaviour with chemical potentials and thermodynamics}",
    eprint = "1412.1464",
    archivePrefix = "arXiv",
    primaryClass = "hep-th",
    reportNumber = "CECS-PHY-14-03",
    doi = "10.1007/JHEP05(2015)025",
    journal = "JHEP",
    volume = "05",
    pages = "025",
    year = "2015"
}

@article{Ammon:2017vwt,
    author = "Ammon, Martin and Grumiller, Daniel and Prohazka, Stefan and Riegler, Max and Wutte, Raphaela",
    title = "{Higher-Spin Flat Space Cosmologies with Soft Hair}",
    eprint = "1703.02594",
    archivePrefix = "arXiv",
    primaryClass = "hep-th",
    reportNumber = "TUW--17--01",
    doi = "10.1007/JHEP05(2017)031",
    journal = "JHEP",
    volume = "05",
    pages = "031",
    year = "2017"
}

@article{Bergshoeff:1989ns,
    author = "Bergshoeff, E. and Blencowe, M.P. and Stelle, K.S.",
    title = "{Area Preserving Diffeomorphisms and Higher Spin Algebra}",
    reportNumber = "IMPERIAL/TH/88-89/9",
    doi = "10.1007/BF02108779",
    journal = "Commun. Math. Phys.",
    volume = "128",
    pages = "213",
    year = "1990"
}

@article{Bekaert:2010hw,
      author         = "Bekaert, Xavier and Boulanger, Nicolas and Sundell, Per",
      title          = "{How higher-spin gravity surpasses the spin two barrier:
                        no-go theorems versus yes-go examples}",
      journal        = "Rev. Mod. Phys.",
      volume         = "84",
      year           = "2012",
      pages          = "987-1009",
      doi            = "10.1103/RevModPhys.84.987",
      eprint         = "1007.0435",
      archivePrefix  = "arXiv",
      primaryClass   = "hep-th",
      SLACcitation   = "%%CITATION = ARXIV:1007.0435;%%"
}

@article{Afshar:2013vka,
      author         = "Afshar, Hamid and Bagchi, Arjun and Fareghbal, Reza and Grumiller, Daniel and Rosseel, Jan",
      title          = "{Spin-3 Gravity in Three-Dimensional Flat Space}",
      journal        = "Phys. Rev. Lett.",
      volume         = "111",
      year           = "2013",
      number         = "12",
      pages          = "121603",
      doi            = "10.1103/PhysRevLett.111.121603",
      eprint         = "1307.4768",
      archivePrefix  = "arXiv",
      primaryClass   = "hep-th",
      reportNumber   = "TUW-13-09",
      SLACcitation   = "%%CITATION = ARXIV:1307.4768;%%"
}

@article{Ammon:2011ua,
      author         = "Ammon, Martin and Kraus, Per and Perlmutter, Eric",
      title          = "{Scalar fields and three-point functions in $D=3$ higher
                        spin gravity}",
      journal        = "JHEP",
      volume         = "07",
      year           = "2012",
      pages          = "113",
      doi            = "10.1007/JHEP07(2012)113",
      eprint         = "1111.3926",
      archivePrefix  = "arXiv",
      primaryClass   = "hep-th",
      SLACcitation   = "%%CITATION = ARXIV:1111.3926;%%"
}

@article{Gaberdiel:2011wb,
    author = "Gaberdiel, Matthias R. and Hartman, Thomas",
    title = "{Symmetries of Holographic Minimal Models}",
    eprint = "1101.2910",
    archivePrefix = "arXiv",
    primaryClass = "hep-th",
    doi = "10.1007/JHEP05(2011)031",
    journal = "JHEP",
    volume = "05",
    pages = "031",
    year = "2011"
}

@article{Grumiller:2014lna,
    author = "Grumiller, D. and Riegler, M. and Rosseel, J.",
    title = "{Unitarity in three-dimensional flat space higher spin theories}",
    eprint = "1403.5297",
    archivePrefix = "arXiv",
    primaryClass = "hep-th",
    reportNumber = "TUW-14-05",
    doi = "10.1007/JHEP07(2014)015",
    journal = "JHEP",
    volume = "07",
    pages = "015",
    year = "2014"
}

@mastersthesis{Riegler:2016hah,
    author = "Riegler, Max",
    title = "{How General Is Holography?}",
    eprint = "1609.02733",
    archivePrefix = "arXiv",
    primaryClass = "hep-th",
    type = "PhD thesis",
    month = "9",
    year = "2016"
}

@article{Campoleoni:2010zq,
    author = "Campoleoni, Andrea and Fredenhagen, Stefan and Pfenninger, Stefan and Theisen, Stefan",
    title = "{Asymptotic symmetries of three-dimensional gravity coupled to higher-spin fields}",
    eprint = "1008.4744",
    archivePrefix = "arXiv",
    primaryClass = "hep-th",
    reportNumber = "AEI-2010-140",
    doi = "10.1007/JHEP11(2010)007",
    journal = "JHEP",
    volume = "11",
    pages = "007",
    year = "2010"
}

@article{Ammon:2011nk,
    author = "Ammon, Martin and Gutperle, Michael and Kraus, Per and Perlmutter, Eric",
    title = "{Spacetime Geometry in Higher Spin Gravity}",
    eprint = "1106.4788",
    archivePrefix = "arXiv",
    primaryClass = "hep-th",
    doi = "10.1007/JHEP10(2011)053",
    journal = "JHEP",
    volume = "10",
    pages = "053",
    year = "2011"
}

@article{Witten:1988hc,
    author = "Witten, Edward",
    title = "{(2+1)-Dimensional Gravity as an Exactly Soluble System}",
    reportNumber = "IASSNS-HEP-88-32",
    doi = "10.1016/0550-3213(88)90143-5",
    journal = "Nucl. Phys. B",
    volume = "311",
    pages = "46",
    year = "1988"
}

@article{PhysRev.135.B1049,
  title = {Photons and Gravitons in $S$-Matrix Theory: Derivation of Charge Conservation and Equality of Gravitational and Inertial Mass},
  author = {Weinberg, Steven},
  journal = {Phys. Rev.},
  volume = {135},
  issue = {4B},
  pages = {B1049--B1056},
  numpages = {0},
  year = {1964},
  month = {8},
  publisher = {American Physical Society},
  doi = {10.1103/PhysRev.135.B1049},
  url = {https://link.aps.org/doi/10.1103/PhysRev.135.B1049}
}

@article{PhysRev.159.1251,
  title = {All Possible Symmetries of the $S$ Matrix},
  author = {Coleman, Sidney and Mandula, Jeffrey},
  journal = {Phys. Rev.},
  volume = {159},
  issue = {5},
  pages = {1251--1256},
  numpages = {0},
  year = {1967},
  month = {7},
  publisher = {American Physical Society},
  doi = {10.1103/PhysRev.159.1251},
  url = {https://link.aps.org/doi/10.1103/PhysRev.159.1251}
}

@article{Gonzalez:2013oaa,
    author = "Gonzalez, Hernan A. and Matulich, Javier and Pino, Miguel and Troncoso, Ricardo",
    title = "{Asymptotically flat spacetimes in three-dimensional higher spin gravity}",
    eprint = "1307.5651",
    archivePrefix = "arXiv",
    primaryClass = "hep-th",
    reportNumber = "CECS-PHY-13-06",
    doi = "10.1007/JHEP09(2013)016",
    journal = "JHEP",
    volume = "09",
    pages = "016",
    year = "2013"
}

@inbook{Prohazka:2017lqb,
    author = "Prohazka, Stefan and Riegler, Max",
    title = "{Higher Spins Without (Anti-)de Sitter}",
    eprint = "1710.11105",
    archivePrefix = "arXiv",
    primaryClass = "hep-th",
    doi = "10.3390/universe4010020",
    volume = "4",
    number = "1",
    pages = "20",
    month = "1",
    year = "2018"
}

@article{Weinberg:1980kq,
    author = "Weinberg, Steven and Witten, Edward",
    title = "{Limits on Massless Particles}",
    reportNumber = "HUTP-80/A056",
    doi = "10.1016/0370-2693(80)90212-9",
    journal = "Phys. Lett. B",
    volume = "96",
    pages = "59--62",
    year = "1980"
}

@article{Barnich:2012aw,
    author = "Barnich, Glenn and Gomberoff, Andres and Gonzalez, Hernan A.",
    title = "{The Flat limit of three dimensional asymptotically anti-de Sitter spacetimes}",
    eprint = "1204.3288",
    archivePrefix = "arXiv",
    primaryClass = "gr-qc",
    doi = "10.1103/PhysRevD.86.024020",
    journal = "Phys. Rev. D",
    volume = "86",
    pages = "024020",
    year = "2012"
}

@article{Ponomarev:2016lrm,
    author = "Ponomarev, Dmitry and Skvortsov, E.D.",
    title = "{Light-Front Higher-Spin Theories in Flat Space}",
    eprint = "1609.04655",
    archivePrefix = "arXiv",
    primaryClass = "hep-th",
    doi = "10.1088/1751-8121/aa56e7",
    journal = "J. Phys. A",
    volume = "50",
    number = "9",
    pages = "095401",
    year = "2017"
}

@article{Skvortsov:2020pnk,
    author = "Skvortsov, Evgeny and Tran, Tung and Tsulaia, Mirian",
    title = "{A Stringy theory in three dimensions and Massive Higher Spins}",
    eprint = "2006.05809",
    archivePrefix = "arXiv",
    primaryClass = "hep-th",
    month = "6",
    year = "2020"
}

@article{Bengtsson:1983pd,
    author = "Bengtsson, Anders K.H. and Bengtsson, Ingemar and Brink, Lars",
    title = "{Cubic Interaction Terms for Arbitrary Spin}",
    reportNumber = "GOTEBORG-83-10",
    doi = "10.1016/0550-3213(83)90140-2",
    journal = "Nucl. Phys. B",
    volume = "227",
    pages = "31--40",
    year = "1983"
}

@article{Bengtsson:1986kh,
    author = "Bengtsson, Anders K.H. and Bengtsson, Ingemar and Linden, Noah",
    title = "{Interacting Higher Spin Gauge Fields on the Light Front}",
    reportNumber = "QMC-86-24",
    doi = "10.1088/0264-9381/4/5/028",
    journal = "Class. Quant. Grav.",
    volume = "4",
    pages = "1333",
    year = "1987"
}

@inproceedings{Metsaev:1996pd,
    author = "Metsaev, R.R.",
    title = "{Cubic interaction vertices for higher spin fields}",
    booktitle = "{2nd International Sakharov Conference on Physics}",
    eprint = "hep-th/9705048",
    archivePrefix = "arXiv",
    pages = "509--514",
    month = "5",
    year = "1996"
}

@article{Metsaev:2020gmb,
    author = "Metsaev, R.R.",
    title = "{Cubic interactions of arbitrary spin fields in 3d flat space}",
    eprint = "2005.12224",
    archivePrefix = "arXiv",
    primaryClass = "hep-th",
    reportNumber = "FIAN-TD-2020-15",
    month = "5",
    year = "2020"
}

@article{Sleight:2016xqq,
    author = "Sleight, Charlotte and Taronna, Massimo",
    title = "{Higher-Spin Algebras, Holography and Flat Space}",
    eprint = "1609.00991",
    archivePrefix = "arXiv",
    primaryClass = "hep-th",
    reportNumber = "MPP-2016-274",
    doi = "10.1007/JHEP02(2017)095",
    journal = "JHEP",
    volume = "02",
    pages = "095",
    year = "2017"
}

@article{Gaberdiel:2012ku,
    author = "Gaberdiel, Matthias R. and Gopakumar, Rajesh",
    title = "{Triality in Minimal Model Holography}",
    eprint = "1205.2472",
    archivePrefix = "arXiv",
    primaryClass = "hep-th",
    reportNumber = "HRI-P-12-05-001",
    doi = "10.1007/JHEP07(2012)127",
    journal = "JHEP",
    volume = "07",
    pages = "127",
    year = "2012"
}

@article{Gaberdiel:2012uj,
    author = "Gaberdiel, Matthias R. and Gopakumar, Rajesh",
    title = "{Minimal Model Holography}",
    eprint = "1207.6697",
    archivePrefix = "arXiv",
    primaryClass = "hep-th",
    doi = "10.1088/1751-8113/46/21/214002",
    journal = "J. Phys. A",
    volume = "46",
    pages = "214002",
    year = "2013"
}

@article{Dirac:1936tg,
    author = "Dirac, Paul A.M.",
    title = "{Relativistic wave equations}",
    doi = "10.1098/rspa.1936.0111",
    journal = "Proc. Roy. Soc. Lond. A",
    volume = "155",
    pages = "447--459",
    year = "1936"
}

@article{Grumiller:2017sjh,
    author = "Grumiller, Daniel and Merbis, Wout and Riegler, Max",
    title = "{Most general flat space boundary conditions in three-dimensional Einstein gravity}",
    eprint = "1704.07419",
    archivePrefix = "arXiv",
    primaryClass = "hep-th",
    reportNumber = "TUW-17-04",
    doi = "10.1088/1361-6382/aa8004",
    journal = "Class. Quant. Grav.",
    volume = "34",
    number = "18",
    pages = "184001",
    year = "2017"
}

@article{Rahman:2015pzl,
    author = "Rahman, Rakibur and Taronna, Massimo",
    title = "{From Higher Spins to Strings: A Primer}",
    eprint = "1512.07932",
    archivePrefix = "arXiv",
    primaryClass = "hep-th",
    month = "12",
    year = "2015"
}

@article{Basu:2015evh,
    author = "Basu, Rudranil and Riegler, Max",
    title = "{Wilson Lines and Holographic Entanglement Entropy in Galilean Conformal Field Theories}",
    eprint = "1511.08662",
    archivePrefix = "arXiv",
    primaryClass = "hep-th",
    doi = "10.1103/PhysRevD.93.045003",
    journal = "Phys. Rev. D",
    volume = "93",
    number = "4",
    pages = "045003",
    year = "2016"
}

@article{Taronna:2011kt,
    author = "Taronna, M.",
    title = "{Higher-Spin Interactions: four-point functions and beyond}",
    eprint = "1107.5843",
    archivePrefix = "arXiv",
    primaryClass = "hep-th",
    doi = "10.1007/JHEP04(2012)029",
    journal = "JHEP",
    volume = "04",
    pages = "029",
    year = "2012"
}

@article{Metsaev:2007rn,
    author = "Metsaev, R. R.",
    title = "{Cubic interaction vertices for fermionic and bosonic arbitrary spin fields}",
    eprint = "0712.3526",
    archivePrefix = "arXiv",
    primaryClass = "hep-th",
    reportNumber = "FIAN-TD-2007-25",
    doi = "10.1016/j.nuclphysb.2012.01.022",
    journal = "Nucl. Phys. B",
    volume = "859",
    pages = "13--69",
    year = "2012"
}

@article{Metsaev:2022yvb,
    author = "Metsaev, R. R.",
    title = "{Interacting massive and massless arbitrary spin fields in 4d flat space}",
    eprint = "2206.13268",
    archivePrefix = "arXiv",
    primaryClass = "hep-th",
    reportNumber = "FIAN-TD-2022-02",
    doi = "10.1016/j.nuclphysb.2022.115978",
    journal = "Nucl. Phys. B",
    volume = "984",
    pages = "115978",
    year = "2022"
}

@article{Campoleoni:2020ejn,
    author = "Campoleoni, Andrea and Francia, Dario and Heissenberg, Carlo",
    title = "{On asymptotic symmetries in higher dimensions for any spin}",
    eprint = "2011.04420",
    archivePrefix = "arXiv",
    primaryClass = "hep-th",
    reportNumber = "NORDITA 2020-103",
    doi = "10.1007/JHEP12(2020)129",
    journal = "JHEP",
    volume = "12",
    pages = "129",
    year = "2020"
}

@article{Gaberdiel:2010pz,
    author = "Gaberdiel, Matthias R. and Gopakumar, Rajesh",
    title = "{An AdS$_{3}$ Dual for Minimal Model CFTs}",
    eprint = "1011.2986",
    archivePrefix = "arXiv",
    primaryClass = "hep-th",
    doi = "10.1103/PhysRevD.83.066007",
    journal = "Phys. Rev. D",
    volume = "83",
    pages = "066007",
    year = "2011"
}

@article{Chang:2011mz,
    author = "Chang, Chi-Ming and Yin, Xi",
    title = "{Higher Spin Gravity with Matter in $AdS_3$ and Its CFT Dual}",
    eprint = "1106.2580",
    archivePrefix = "arXiv",
    primaryClass = "hep-th",
    doi = "10.1007/JHEP10(2012)024",
    journal = "JHEP",
    volume = "10",
    pages = "024",
    year = "2012"
}

@article{Rahman:2012thy,
    author = "Rahman, Rakibur",
    title = "{Higher Spin Theory - Part I}",
    eprint = "1307.3199",
    archivePrefix = "arXiv",
    primaryClass = "hep-th",
    doi = "10.22323/1.195.0004",
    journal = "PoS",
    volume = "ModaveVIII",
    pages = "004",
    year = "2012"
}

@article{Prohazka:2017equ,
    author = {Prohazka, Stefan and Salzer, Jakob and Sch\"oller, Friedrich},
    title = "{Linking Past and Future Null Infinity in Three Dimensions}",
    eprint = "1701.06573",
    archivePrefix = "arXiv",
    primaryClass = "hep-th",
    doi = "10.1103/PhysRevD.95.086011",
    journal = "Phys. Rev. D",
    volume = "95",
    number = "8",
    pages = "086011",
    year = "2017"
}

@article{Blencowe:1988gj,
    author = "Blencowe, M. P.",
    title = "{A Consistent Interacting Massless Higher Spin Field Theory in $D$ = (2+1)}",
    reportNumber = "IMPERIAL/TP-87-88/30",
    doi = "10.1088/0264-9381/6/4/005",
    journal = "Class. Quant. Grav.",
    volume = "6",
    pages = "443",
    year = "1989"
}

@article{Gutperle:2011kf,
    author = "Gutperle, Michael and Kraus, Per",
    title = "{Higher Spin Black Holes}",
    eprint = "1103.4304",
    archivePrefix = "arXiv",
    primaryClass = "hep-th",
    doi = "10.1007/JHEP05(2011)022",
    journal = "JHEP",
    volume = "05",
    pages = "022",
    year = "2011"
}

@article{Bunster:2014mua,
    author = "Bunster, Claudio and Henneaux, Marc and Perez, Alfredo and Tempo, David and Troncoso, Ricardo",
    title = "{Generalized Black Holes in Three-dimensional Spacetime}",
    eprint = "1404.3305",
    archivePrefix = "arXiv",
    primaryClass = "hep-th",
    doi = "10.1007/JHEP05(2014)031",
    journal = "JHEP",
    volume = "05",
    pages = "031",
    year = "2014"
}

@article{Bekaert:2022ipg,
    author = "Bekaert, Xavier and Oblak, Blagoje",
    title = "{Massless scalars and higher-spin BMS in any dimension}",
    eprint = "2209.02253",
    archivePrefix = "arXiv",
    primaryClass = "hep-th",
    doi = "10.1007/JHEP11(2022)022",
    journal = "JHEP",
    volume = "11",
    pages = "022",
    year = "2022"
}
\end{document}